\documentclass[a4paper,fleqn]{cas-sc}

\usepackage[numbers]{natbib}

\usepackage{graphicx}
\usepackage[table]{xcolor}
\usepackage{multirow}
\usepackage{longtable}
\usepackage{tabularx,booktabs}

\usepackage{amsmath,amsfonts}
\usepackage{array}
\usepackage[caption=false,font=normalsize,labelfont=sf,textfont=sf]{subfig}
\usepackage{textcomp}
\usepackage{stfloats}
\usepackage{url}
\usepackage{verbatim}
\usepackage{graphicx}
\usepackage{cite}
\usepackage{amsmath}             
\usepackage{mathtools}
\usepackage{mathdots}    
\usepackage{amssymb}   
\usepackage{subcaption}
\usepackage{algpseudocode}

\usepackage{graphicx}
\usepackage[table]{xcolor}
\usepackage{multirow}
\usepackage{longtable}
\usepackage{tabularx,booktabs}

\usepackage{array}

\usepackage{graphicx}
\usepackage{subcaption} 

\usepackage[ruled,vlined,linesnumbered]{algorithm2e}

\usepackage{amssymb}

\usepackage{multirow}

\usepackage{placeins}
\def\tsc#1{\csdef{#1}{\textsc{\lowercase{#1}}\xspace}}
\tsc{WGM}
\tsc{QE}

\begin{document}
\let\WriteBookmarks\relax
\def\floatpagepagefraction{1}
\def\textpagefraction{.001}

\shorttitle{}    

\shortauthors{}  

\title [mode = title]{EVNextTrade: Learning-to-Rank-Based Recommendation of Next Charging Nodes for EV--EV Energy Trading}  



%















\author[1]{Md Mahfujur Rahman}[orcid=0000-0002-3102-2099]
\cormark[1]
\ead{mdmahfujur.rahman@hdr.qut.edu.au}

\author[1]{Alistair Barros}[orcid=0000-0001-8980-6841]

\author[2]{Raja Jurdak}[orcid=0000-0001-7517-0782]

\author[1]{Darshika Koggalahewa}[orcid=0000-0002-2695-1846]

\affiliation[1]{organization={School of Information Systems, Queensland University of Technology},
            city={Brisbane},
            state={QLD},
            postcode={4000},
            country={Australia}}

\affiliation[2]{organization={School of Computer Science, Queensland University of Technology},
            city={Brisbane},
            state={QLD},
            postcode={4000},
            country={Australia}}

\cortext[1]{Corresponding author}
\fntext[1]{}


\begin{abstract}
Peer-to-peer energy trading among electric vehicles (EVs) has been increasingly studied as a promising solution for improving supply-side resilience under growing charging demand and constrained charging infrastructure. While prior studies on EV--EV energy trading and related EV research have largely focused on transaction management or isolated mobility prediction tasks, the problem of identifying which charging nodes are more suitable for EV--EV trading
in journey contexts remains open. We address this gap by formulating next charging nodes recommendation as a learning-to-rank problem, where each EV decision event is associated with a set of candidate charging locations. We propose a supervised ranking framework applied to a large-scale urban EV mobility dataset comprising millions of journey records and multidimensional EV trading-related features, including EV energy level, trading role, distance to charging locations, charging speed, and temporal station popularity. To account for uncertainty arising from the mobility of both energy providers and consumers, as well as the presence of multiple viable charging nodes at a decision point, we employ probabilistic relevance refinement to generate graded labels for ranking. We evaluate gradient-boosted learning-to-rank models, including LightGBM, XGBoost, and CatBoost, on EV journey records enriched with candidate charging nodes. Experimental results show that LightGBM consistently achieves the strongest ranking performance across standard metrics, including NDCG@k, Recall@k, and MRR, with particularly strong early-ranking quality, reflected in the highest NDCG@1 (0.9795) and MRR (0.9990). These results highlight the effectiveness of uncertainty-aware learning-to-rank for charging node recommendation and support improved coordination and matching in decentralized EV--EV energy trading systems.
\end{abstract}


\maketitle


\section{Introduction}

Electric vehicles (EVs) have gained significant attention due to their potential to offer sustainable and eco-friendly transportation solutions. The growing popularity of EVs, the declining cost of Renewable Energy Resources (RES), and technological advancements in energy storage are significant factors in this trend. By 2030, the number of EVs on the road worldwide is expected to increase from just over 11 million in 2020 to approximately 145 million, according to the International Energy Agency (IEA). At the same time, the increasing number of EVs has led to new and persistent challenges. Charging demand for EVs is surging and faces barriers due to the limited number of available grid-connected charging stations \citep{P1-8}. Meanwhile, EV drivers face bottlenecks, queues, and prolonged wait times when accessing charging stations (CSs). The limited availability of grid-connected CSs exacerbates this challenge \citep{P316-1}. Added to these challenges are localized energy deficits, where grid constraints and insufficient charging infrastructure have already limited EV usage, leading to a decline in EV adoption despite increasing mobility demand \citep{P1}. Finally, as widely reported, EV drivers experience range anxiety due to the limitations of charge stations along their travel routes, which are often commensurate with their daily journeys, and the limited battery lifetimes of EVs, requiring frequent recharging.

In the face of these challenges, new opportunities are emerging to address resource shortages in energy supply and distribution beyond the central urban areas in which EV energy trading is prominent \citep{C62, rahman2025survey}. In particular, bidirectional energy exchange across EVs is growing in maturity, as is the reliability of wireless charging pads (WCPs) through which EV--EV trading can take place \citep{P1, C58, P15}. However, the contexts for EV--EV trading carry key differences compared to the currently available EV--to--CS (EV--CS) models. Consumers, as well as providers, need to be matched with proximate EVs, each carrying distinct and independent mobile contexts. In particular, the selection of trade points involves navigation through varying travel pathways, each consisting of different road segments with variable traffic densities. This usually results in variations in driver journey routes, leading to choices and decisions about the selection points for trading. In addition, the choices are influenced by the perceived waiting times at charging nodes, which include the times of energy transfer and queuing times as cars await their turn for access to charging nodes. These challenges are further heightened, as successful bilateral trades depend on the alignment of independent drivers’ navigation choices, each navigating their own journey and diminishing energy levels. Ultimately, a trade, being essentially a service, carries delivery obligations of resource capacity (how much energy is expected to be exchanged and the speed of transfer) and reliability (the availability of the trading partner). This, too, needs to be carefully appraised by drivers when making choices about where to trade energy.

This highlights the need for potential energy management solutions.  
Existing research in EV energy management has made significant progress in forecasting charging demand and analyzing EV mobility, particularly for stationary grid-connected charging stations. Machine learning and deep learning techniques have been widely applied to predict charging demand and infer charging behavior from historical data \citep{RQ1-C1,P3}. However, most prior studies assume relatively stable charging availability and largely focus on demand-side behavior, limiting its applicability to EV–EV energy trading scenarios where both energy providers and consumers are mobile \citep{P316-1}. While some studies classify EVs as energy providers or consumers based on predicted energy consumption \citep{P294}, they typically rely on limited travel history and overlook critical factors such as distance to candidate charging locations, charging-site capacity, charging speed, and the presence of nearby trading partners. Similarly, mobility-based approaches that predict next charging location assume guaranteed charging availability at predicted vehicle destination, an assumption that is frequently unrealistic in EV–EV trading contexts \citep{C57}.  

From a data perspective, existing EV datasets research predominantly capture charging-session records at fixed stations \citep{link-1,workplacedata,acndata}, or general mobility traces without explicitly modeling of energy trading interactions \citep{mobilitydata}. As a result, there is a lack of integrated, multidimensional datasets that jointly model EV journeys, dynamic trading roles, energy exchange requirements, and candidate charging locations along with their associated attributes.

Crucially, EV–EV energy trading introduces a complex decision-support problem that exceeds the scope of existing prediction-based approaches. An EV–EV trading requires the spatio-temporal synchronization of two mobile EVs at a specific node.
This interaction is parameterized by multiple competing variables, including travel detours, waiting times, energy availability, charging speeds, and the availability of successful partner matching. Furthermore, each EV operates within a unique mobile context, defined by distinct journey requirements and navigation choices across road networks with varying traffic densities. As such, EV drivers require automated support to coordinate journeys and decide where to consume or provide energy. These limitations motivate the need for recommendation models tailored to EV–EV energy trading, which jointly consider EV mobility, dynamic energy state, charging infrastructure characteristics, and contextual factors from both the energy provider and consumer perspectives. In this study, we address this gap by formulating next charging node recommendation as a ranking problem, in which each EV decision event is associated with a set of candidate charging nodes that are evaluated and ranked simultaneously. Rather than predicting a single future location, the objective is to rank candidate charging nodes according to their relative likelihood of being selected for energy trading at a given decision point along an EV journey.

To support next charging node recommendation in EV–EV energy trading, we propose a supervised ranking-based framework that integrates mobility and energy-related features including EV battery state, trading role, distance to candidate charging locations, charging speed, and temporal station popularity. The framework produces an event-specific (i.e., for each EV, journey, and time) ordering of candidate charging nodes based on EV journeys, battery energy levels, distances to charging locations, charging speeds, and station popularity. Experimental results demonstrate the effectiveness of the proposed approach in identifying the most likely next charging nodes for EV--EV energy trading.
This paper contributes to the advancement of the field through the following key contributions:

\begin{itemize}
    \item We present a multidimensional EV–EV trading dataset that includes EV journeys, candidate charging nodes (WCPs), and EV trading intention profiles for next charging nodes recommendation. The dataset further incorporates charging station metadata, including location, charger type, charging speed, and number of ports, as well as Google Popular Times data to capture temporal user density and energy availability trends at WCPs.

    \item To the best of our knowledge, this is the first study to address the next charging nodes recommendation specifically for EV–EV energy trading. We formulate the problem as a ranking task and propose a learning-to-rank framework that leverages multidimensional EV mobility and energy-related data to predict the relative likelihood of charging nodes being considered for EV–EV energy trading at a given decision point along an EV journey. 

     \item We evaluate multiple supervised learning-to-rank models, including LightGBM, XGBoost, and CatBoost, under identical experimental settings. The results demonstrate robust and consistent ranking performance across models, highlighting the effectiveness of the proposed framework in learning relative charging node preferences under uncertainty.
  
\end{itemize}

The paper is structured as follows. Section \ref{rw} reviews the related work on EV charging supply and demand prediction, EV mobility prediction, charging location recommendation systems for EVs, and features used for mobility prediction, and identifies the corresponding research issues. Section \ref{rfsr} presents the rationale for future-state recommendation. Section \ref{sd} describes the study design, and Section \ref{pf} formulates the problem. Section \ref{rm} details the research methodology, including data acquisition, label generation, data preprocessing, learning-to-rank modeling, model training, and evaluation metrics. Section \ref{rad} presents the experimental results and discussion. Section \ref{limitation} outlines the limitations of the study, and Section \ref{conclusion} concludes the paper and discusses future research directions.

\section{Related work}
\label{rw}
This section reviews developments and key contributions to the existing literature on predictive and recommendation techniques, either directly related to EV--EV trading or indirectly related through proposals from EV-CS. 
First, it discusses several techniques for predicting charging demand and supply for EV–EV and EV--CS settings, highlighting their relevance to EV–EV trading. Next, it discusses techniques focusing on EV mobility prediction. Subsequently, it discusses a charging location recommendation system for EVs. Finally, it identifies the salient features used in mobility prediction and delves into the research issues that need to be addressed to advance the EV–EV energy trading systems.

\subsection{EV charging supply and demand prediction}
Accurate prediction of charging demand and supply is essential for matching energy providers and consumers in EV--EV trading. Zhao et al. \citep{P294} used a LightGBM classifier to identify providers and consumers based on expected mileage and real-time location, achieving 95\% accuracy in role classification. However, their model excluded contextual mobility features such as journey navigation choices of the driver, traffic congestion, and detour distances, limiting spatial-temporal applicability. Yi et al. \citep{P3} applied a Seq2Seq deep learning model for long-term charging demand forecasting at fixed charging stations (CSs), outperforming ARIMA, LSTM, and XGBoost baselines, but without addressing supply-side variability. Similarly, by integrating heterogeneous features like historical charging data and time-related factors (e.g., time of day, day of the week), the T-LSTM-Enc \citep{P7} and hybrid LSTM network \citep{P343} algorithms can predict energy demand at CSs, capturing long-term and short-term features of charging demand. However, they did not incorporate real-time supply or mobility data. These studies indicated the importance of heterogeneous features in developing effective coordination strategies for improved trading matching and scheduling at charging locations. In addition, these studies also showed that accurate prediction enables charging station operators to optimize resource allocation and manage energy supply more effectively. 

\subsection{EV mobility prediction}

This section explores EV mobility prediction, where EVs are candidates for providing and consuming energy in trading. To facilitate EV--EV trading, EVs need to be effectively redirected to convenient charging locations where they can provide or consume energy. However, a pair of EV providers and consumers is required for trading, with the challenge of EVs being mobile on different routes and in different geographic areas. Hence, this section covers the existing literature on individual mobility predictions of EVs and investigates the current state of research in this domain.

Over the past few years, DL-based neural networks have been increasingly adopted for individual vehicle location prediction. The LSTM model with a backtracking attention mechanism addresses long-term information handling in sparse scenarios and mitigates vanishing sequential dependency in vehicle stop location prediction \citep{mob3}. This mechanism aggregates historical hidden states using weights that reflect user regularity and preference, improving prediction accuracy by 7-10\% over state-of-the-art methods across three real-world datasets with over 10,000 vehicles. The study \citep{mob15} also used LSTM to predict the next destinations of taxi drivers based on sequences of pick-up and drop-off points and timestamps. They utilized bag-of-concept (BOC) and word2vec methods to capture points of interest (POIs), such as shopping malls and schools, and spatial zone embedding for accurate predictions. Herberth et al. \citep{traffic-mob-2} also mentioned POIs as significant parameters for mobility prediction. 

Wang et al. \citep{mob21} proposed a hybrid model that integrates a multi-layer perception (MLP) to extract local features and LSTM to extract long-term dependency for predicting the trajectory of only the moving state of taxis \citep{mob21}. The results showed that the MAE increased when the number of future prediction steps was long. Similarly, Fan et al. \citep{mob10} introduced a hybrid model combining CNN to capture local features and bidirectional LSTM to capture global features. This model predicts the next location in each trajectory by considering contextual information like traffic flow and weather data. The results showed that the model's accuracy surpasses the baseline methods, particularly in predicting longer sequences of future locations.

Although these studies do not directly address EV mobility prediction, their relevant methodologies can provide useful insights for developing EV mobility prediction at WCPs. Some studies have been attempted on the prediction of EV driving range \citep{C37}, EV next charge location \citep{C57}, etc., which may also be the pioneering idea for EV mobility prediction in EV--EV trading.  
A study by Robert et al. \citep{C57} used a CNN to predict EVs' next charging locations based on battery levels, providing a novel dataset for EV mobility. 
However, this study focused only on demand-side prediction, overlooking energy trading, which requires predicting the mobility of both energy providers and consumers at charging locations. 
Additionally, this study did not account for distances to the charge locations and charging speed, which could impact predictions. 

Based on these studies, there is a lack of EV mobility data at WCPs with varying energy requirements and their ability to provide and obtain energy. However, DL-based techniques such as CNN, LSTM, and CNN-LSTM are promising solutions for solving EV mobility prediction and learning long-term dependencies in mobility time-series data.

\subsection{Charging location recommendation system for EVs}

Teimoori et al. [12] presented an EV recommender model for charging station selection based on a combination of cloud and federated learning. The selection process is done based on some variables such as the availability of EVCS charging spots, distance, remaining battery capacity, price, and time. The weights are calculated by dividing the parameters into EV and EVCS-related parameters and forming a vector of pairs representing the variables involved in the computation. The average of all EVCS variable values is assigned as a weight for that variable. Since the proposed model is assumed to recommend an EVCS in real-time, the latency can be notified in the selection process because of insufficient EV and EVCS sample size, which can affect the execution time and, consequently, the performance of the recommender system. However, the supply side has not been taken into consideration while selecting the best EVCS. Another study by Habbal et al. \citep{habbal2024user} developed a User-Preference-Based Charging Station Recommendation Scheme (UPCSRS) that integrates user preferences into a multiple-attribute decision-making (MADM) framework. The approach employs the analytic hierarchy process (AHP) to assign weights to user-defined criteria--distance, waiting time, charging time, and price--and applies the technique for order preference by similarity to the ideal solution (TOPSIS) to rank the available stations. Using real data from the U.S. Department of Energy, the scheme achieved superior accuracy and user satisfaction compared with proximity- or price-based methods, demonstrating the effectiveness of a user-centric and adaptive recommendation process for EV–station selection. However, their model primarily focuses on EV–station interactions and does not consider decentralized energy exchange between vehicles.

A large number of selection criteria for charging locations can cause a decline in system performance, while a smaller number may not result in a good decision \citep{habbal2024user}. Therefore, to determine the appropriate range of
criteria, a review was conducted, ranging from three \citep{habbal2024user, tian2016real, wang2025beyond, dixon2018characterization, dixon2019evaluating, mahdi2023modeling, hossain2025cloud} to ten. However, according to the literature review, distance, charging time (calculated by charging speed), Google popular times, waiting time, and price emerged as key determinants from a comprehensive analysis of prior studies in the field. 

Beyond EV charging, recommendation research in related domains, such as point-of-interest (POI) selection in location-based services and e-commerce, has adopted a range of methodologies, including fuzzy logic, ranking-based approaches, and probabilistic modeling \citep{mao2015fuzzy, zehlike2022fairness, moayedikia2026trust}. Fuzzy content matching and fuzzy multi-criteria decision-making methods are particularly effective in settings where user preferences and contextual factors are uncertain, as they model relevance in a graded rather than binary manner. In particular, Fuzzy TOPSIS has been widely used to aggregate heterogeneous criteria and produce ordered candidate lists when relevance is imprecise or derived indirectly from multiple factors \citep{sodhi2012simplified, mieszkowicz2020preference}.

Learning-to-rank has consequently emerged as a principal approach in information retrieval and recommender systems, where the objective is to order candidate items by their relative relevance rather than to predict absolute utility values \citep{zehlike2022fairness, sun2024multi}. In learning-to-rank formulations, each recommendation event is treated as a query over multiple candidates, and models learn preference relationships using pair-wise, or list-wise supervision. Ranking-based formulations align naturally with top-k evaluation metrics and graded relevance modeling, and are well-suited to recommendation scenarios where decisions are comparative.

A persistent challenge in ranking-based recommendation systems is the lack of explicit ground-truth relevance labels. In many real-world settings, such as mobility and location recommendation, relevance is not explicitly provided and must be inferred from noisy, incomplete information. In such cases, relevance signals are often constructed indirectly from available criteria rather than obtained as direct annotations. Fuzzy multi-criteria decision-making methods provide a systematic way to generate graded relevance signals that preserve ordinal relationships among alternatives without imposing hard decisions \citep{sodhi2012simplified, mao2015fuzzy}.

Separately, latent-variable probabilistic models have been widely used in recommender systems to capture hidden user preferences and contextual relevance under sparse or noisy observations, with model parameters typically inferred using the Expectation-Maximization (EM) algorithm \citep{moayedikia2026trust}. These probabilistic approaches focus on uncovering latent relevance structure from noisy, indirectly constructed relevance signals rather than relying on explicit ground-truth supervision, making them well-suited to recommendation scenarios with limited or uncertain feedback.

However, none of these analyses aims to develop a recommendation system for EV--EV energy trading, which involves both demand and supply sides, including EV journeys at wireless charging pads. 

\subsection{Features used for mobility prediction}\label{pred_features}

Features play a critical role in prediction for EVs \citep{P8-17}. 
The accuracy and effectiveness of the predictive models heavily rely on the relevant and informative features extracted from the data. Similarly, the characterization of mobility features, measurement of individual travel behaviors, and performance analysis are crucial for vehicle mobility prediction \citep{mob3}. 
The table \ref{tab:data-mobility} summarizes key features for mobility prediction, including static attributes like vehicle ID and model name, and dynamic attributes such as timestamps, locations, visit frequency, and battery energy levels. These features collectively enable tracking and analyzing vehicle movements, travel patterns, and frequent POIs. Dynamic attributes like battery energy levels and trip distance are particularly relevant attributes in relation to EV--EV trading. 
However, the table underscores that most studies focus on conventional vehicles or individual EV mobility attributes rather than EV--EV trading aspects, leaving a gap in leveraging EV mobility and energy-centric attributes for EV--EV energy trading. This highlights the need for further research integrating EV mobility and energy-centric attributes to address specific challenges in EV--EV trading.

\begin{table*}[!ht]
\centering
\footnotesize
\caption{Common data attributes or features used in existing studies for supporting mobility prediction: mobility data description}
 
 \label{tab:data-mobility}
 \begin{tabularx}{\textwidth}{p{22mm}p{15mm}p{18mm}p{25mm}p{25mm}p{28mm}}

 \hline Field/Features & Category & Static/dynamic & Description & Ref & Used in EV--EV trading\\ 

 \hline

VId & Identifier & S & Unique identifier for the conventional vehicle & \citep{mob3, mob18, mob16, mob15, traffic-mob-2, mob21, mob10} & Yes \\
 \hline
Vehicle model & Vehicle & S & Name of the vehicle model & \citep{C57} & Yes\\
\hline
Timestamps & Temporal & D &  Current date and time & \citep{mob3, C57, mob18, mob16, mob15, traffic-mob-2, mob21, mob10} & Yes\\
 \hline

Locations (latitude, longitude) & Spatial & D & Locations of vehicle at different time  & \citep{mob3, C57, mob18, mob16, mob15, traffic-mob-2, mob21, mob10} & Yes\\
 \hline

Stay locations & Spatial & D & Vehicles stay location at different stoppage & \citep{mob3} & No\\
\hline

Visit frequency & Behavioral & D & Indicate frequent visit at specific locations  & \citep{mob3, mob18, mob3-6} & No\\
 \hline

POIs & Contextual & D & Represents nearby points of interest & \citep{mob15, traffic-mob-2} & No \\
 \hline


Radius of gyration & Mobility & D & Distanced travel by users & \citep{mob22} & No \\
\hline
Battery energy level & Energy & D & State of the charge (SoC) of EVs & \citep{C57} & Yes\\
\hline
Trip distance (km) & Mobility & D & Represent the distance covered in km & \citep{C57} & No\\

\hline

\end{tabularx}
\end{table*}

The studies discussed in this section address similar problem domains using different predictive and recommendation approaches. Their findings have informed the development of our proposed solution by clarifying the existing methodological gaps. Overall, the reviewed research demonstrates substantial progress in EV--CS prediction and recommendation systems but reveals a lack of predictive modeling that captures EV--EV interactions at wireless charging pads. This gap motivates our proposed approach, which focuses on recommending the next charging locations for decentralized EV–EV energy trading.

\subsection{Research issues}

A significant challenge in EV--EV energy trading is anticipating EV mobility and charging node selection while travelling. This difficulty arises from heterogeneous travel patterns, dynamic energy states, varying charging or discharging intent, and the mobility of both energy consumers and providers. These uncertainties make it challenging to determine where EVs are likely to be at any given time for energy trading, particularly when multiple feasible charging locations are available along an EV’s journey.

However, existing studies largely overlook EV mobility contexts that are critical for EV--EV energy trading, such as dynamically changing energy levels, dual supply--demand roles of EVs, start and stop-off locations, destination-dependent routing, proximity to charging resources, and constraints on the capacity and timeliness of energy transfer. More importantly, prior work often treats charging node selection as an isolated problem, without explicitly modeling the relative desirability of multiple feasible charging locations encountered along a journey.

As a result, next charging nodes recommendation in EV--EV energy trading remains underexplored. In practical trading scenarios, EVs must evaluate and choose among a set of reachable wireless charging pads at a given time, and these decisions are inherently relative rather than absolute. For example, a wireless charging pad $WCP_1$ may be considered more suitable than $WCP_2,\ldots, WCP_n$ at time $t$ due to differences in distance, energy transfer capability, availability, and the likelihood of successful interaction with another EV.

This study addresses this gap by developing a predictive framework for next charging node recommendation that ranks candidate charging locations according to their relative suitability for EV--EV energy trading. By explicitly modeling relative desirability under uncertainty, the proposed approach provides a foundation for effective coordination and matching in decentralized EV--EV energy trading systems.

\section{Rationale for future state recommendation}
\label{rfsr}

Our objective is to infer likely EV--EV energy trading locations \emph{before} the next charging cycle occurs, using only historical and contextual information available at a given time. This is motivated by the following considerations:

\begin{enumerate}

    \item \textit{Efficient matching and proactive coordination:} 
    Anticipating future trading locations enables proactive matching between provider and consumer EVs and advanced coordination of energy exchanges. By reasoning about likely future states, the system can mitigate localized congestion at wireless charging pads and reduce coordination delays, thereby improving overall energy distribution efficiency.

    \item \textit{Anticipatory decision modeling:} 
    Prior work in transportation and EV charging behavior suggests that charging-related decisions are often made ahead of arrival based on journey progress, remaining energy, and anticipated charging opportunities. Modeling these decisions in advance supports proactive guidance and aligns with how vehicles plan subsequent actions during a trip.

    \item \textit{Proactive driver support:} 
    The proposed framework targets proactive assistance, such as recommending suitable trading nodes or scheduling energy exchanges ahead of time. This requires predicting likely future charging locations prior to the next charging cycle, rather than reacting only after a vehicle arrives at a charging node.

    \item \textit{Feasibility from observed mobility regularities:} Although the mobility data are derived from taxi trajectories, they exhibit stable spatio-temporal regularities under real-world demand (e.g., recurrent routes, frequent stop locations, and time-of-day concentration around popular destinations). These repeated patterns provide sufficient signal to learn a mapping from historical and contextual features to \emph{likely next} charging/trading locations. The approach does not assume optimal or fully rational behavior; it exploits empirically observed collective regularities to support context-dependent charging-node recommendations.

\end{enumerate}

This approach enables context-aware recommendations that align with how drivers plan their subsequent actions while traveling.

\section{Study design} 
\label{sd}

The aim of this study is to develop techniques and algorithms to support efficient EV--EV trading in which optimal pairs of suppliers and consumers can be matched at locations and times, given dynamic mobile contexts of EVs. 
To support this, this study proposes to study an enriched context for EV mobility and to develop techniques and algorithms in support of a predictive model. This study is guided by one main research question (RQ): \textit{How can the charging of EVs be recommended at nodes (wireless charging pads) for both providers and consumers, given EV mobility contexts?}

By understanding the challenges and research gaps, this question aims to recommend the most likely wireless charging pads for each EV. This RQ will guide the investigation, helping to explore various aspects of predictions and to develop an effective solution for EV--EV trading. To support this, our contextual assumptions are as follows:

\begin{itemize}
    
    \item EV journeys are not strictly point-to-point but may include intermediate stop-offs due to travel or charging needs, reflecting realistic mobility patterns and enabling the consideration of multiple candidate charging or trading nodes along a journey.

    \item A wireless charging pad considers energy sourcing only from the provider.

    \item Wireless charging pads are unevenly distributed in spatial areas.
    
    \item Providers and consumers are unevenly distributed in unbounded spatial areas.

    \item EV driver choice for node selection are based on variable traffic flows, capacity, and timeliness of energy transfer of charging pads, proximity to charging resources, and energy availability.
    
   \item EV drivers will deviate (detour) from their journey routes (e.g., home to work, work to home) by up to 10km only. This is based on travel time, traffic conditions, and energy constraints.

    \item EV drivers will not supply more energy than they require to reach their own destinations.
    
     \item EV drivers will be willingness to participate in EV–EV energy trading when contextual conditions, such as energy availability, travel feasibility, and charging node accessibility, are sufficiently favorable.

    \item The charging amount for supplying and consuming will be sufficient to fulfill the next journey requirement of both provider and consumer for a given route.

    \item Routing decisions made by EV drivers are influenced by navigation applications (e.g., Google Maps or similar services), which guides route based on factors such as travel time, traffic conditions, and distance. This assumption reflects realistic navigation-assisted driving behavior and effects the set of feasible routes and charging nodes encountered during a journey.

    \item Trading system can track the energy used by EVs.

\end{itemize}

These assumptions will guide the formulation of the mobility prediction function.

\section{Problem formulation}
\label{pf}

To support efficient EV–EV energy trading, we formulate a ranking-based predictive model that estimates the relative likelihood of candidate charging nodes for mobile EV providers and consumers at each decision point, given their energy state and mobility context. 
For a given spatial area $sa_i \in \mathcal{SA}$ and decision time $t$, let $\mathcal{E}_t = \{ e_1, e_2, e_3, \ldots \}$
denote the set of EVs located within, or traversing through, $sa_i$ during the decision interval $[t,\, t+\Delta t)$, where $\Delta t = 30$ minutes. 
Each EV $e_i \in \mathcal{E}_t$ is defined by the tuple
$\langle e_{id}, O_{e_i}^t, D_{e_i}^t, l_{e_i}^t, E_i^t, B_c^{e_i}, u_{e_i}^t, d_{e_i}^t \rangle$, corresponding to its journey state $J_{e_i}^t$ at time $t$. Here, $e_{id}$ is the unique identifier of the EV. 
$O_{e_i}^t$ and $D_{e_i}^t$ denote the origin and destination of $e_i$ at time $t$, drawn from the global origin and destination sets
$\mathcal{O}_t = \{ O_{e}^t : e \in \mathcal{E}_t \}, \mathcal{D}_t = \{ D_{e}^t : e \in \mathcal{E}_t \}$.
The variable $l_{e_i}^t \in \mathcal{L}$ represents the spatial location of $e_i$ at time $t$, where $\mathcal{L}$ denotes the set of feasible geographic locations. Locations may correspond to origins, destinations, or intermediate travel or trading points. $E_i^t$ denotes the current battery energy level of $e_i$ at time $t$, expressed as a percentage, and $B_c^{e_i}$ is the battery capacity. The term $u_{e_i}^t$ represents the cumulative energy consumption (Wh) up to time $t$, while $d_{e_i}^t$ denotes the traveled distance within the journey.

Let each EV $e_i$ have a battery capacity $B_c$ (Wh) and a state of charge $\mathrm{SoC}_{e_i}(t) \in [0,100]$ at time $t$. This $SoC$ is an important parameter in EV charging and discharging, as it represents the current charge level and plays a significant role in determining the charging or discharging activities during a journey. Along an EV’s journey, charging and discharging decisions arise at discrete decision points, such as intermediate stop-offs, route deviations, or proximity to charging infrastructure. At such points, an EV may act either as an energy consumer or provider, depending on its available battery energy and remaining travel requirements.

We assume that an EV becomes eligible to act as an energy provider when it possesses sufficient surplus energy beyond its immediate mobility needs. Specifically, if the state of charge exceeds a predefined safety threshold $\mathrm{SoC}_{\mathrm{th}}$, the EV is considered capable of discharging energy without compromising its ability to complete the remaining journey. Following a prior study \citep{P201}, we set $\mathrm{SoC}_{\mathrm{th}} = 30\%$ of the battery capacity $B_c$ for provider EVs $e_i$. The threshold is calculated in Wh using the following equation.

\begin{equation} \label{eq:socchecking}
 E_{th} = \frac{SoC_{th}}{100} * B_c^{e_i} 
\end{equation}

We calculate the current energy $E_i(t)$ in Wh by determining the current SoC level and subtracting $E_{th}$ from $E_i(t)$ to get the available energy $E^{\mathrm{av}}_{i}(t)$.

\begin{equation}
    E_i(t) = \frac{\mathrm{SoC}_{e_i}(t)}{100} \times B_c^{e_i}.
\end{equation}

\begin{equation}\label{eq:avail}
E^{\mathrm{av}}_{i}(t) = E_i(t) - E_{\mathrm{th}},
\end{equation}

If $E^{\mathrm{av}}_{i}(t)$ of an EV $e_i$ is more than minimum tradable energy $(E_{\min}^{\mathrm{trade}})$ which is ten thousand units of energy \citep{P201}, and it has enough energy  $E_i^{\mathrm{travel}}(t,d)$ for going to the next destination \citep{P294}, it can provide energy at proximate charging pads at time t. 

\begin{equation} \label{eq:avail_prov}
     E_i^{\mathrm{prov}}(t) = \max\!\left( 0,\; E^{\mathrm{av}}_{i}(t) -E_i^{\mathrm{travel}}(t,d) \right)
\end{equation}
We assume that EVs can provide energy when they have energy surpluses and the right conditions, such as minimal time to reach charging pads and fast energy transfer. Otherwise, Otherwise, we classify it as a consumer that requires energy at charging pads. 
For EVs acting as energy consumers, we define a upper state-of-charge level $\mathrm{SoC}_{\mathrm{target}}$ (e.g., $100\%$), which represents the preferred charging level for continued travel reliability and charging requirement $\mathrm{SoC}_{\mathrm{min}}$ (e.g., $20\%$), to avoid battery deterioration.
Consumer energy deficit can be expressed as follows:
\begin{equation}\label{cons}
 E_i^{\mathrm{cons}}(t) = 
    \max\!\left( 0,\;
    \frac{\mathrm{SoC}_{\mathrm{target}}}{100} B_i -
    E_i(t) \right)   
\end{equation}

We consider the maximum required distance from the intended route to nodes of each EV $e_i$ for charging or discharging between the source and destination. At each decision time $t$ between source and destination, an EV $e_i$ encounters a set of feasible charging nodes (WCPs), \[
\mathcal{CP}_t^{e_i} = \{CP_1, CP_2, \ldots\},
\]
identified based on distance and accessibility from its current route. Each candidate charging node $CP_j$ is characterized by attributes such as distance to the EV $d(CP_j, e_i(t))$, charging or discharging speed $s(CP_j)$, and time-dependent popularity or congestion level $p(CP_j, t)$ (e.g., derived from Google Popular Times) to understand provider or consumer density of the charging node.

EV drivers are assumed to have access to such information, including charging-node availability as well as aggregate supply and demand conditions through trading applications. These signals influence both the decision to act as an energy provider or consumer and the selection of viable charging nodes at decision time $t$.

Let $\mathbf{x}_{i,t}^{\text{context}}$ denote the feature vector describing the EV journey, energy level, spatial-temporal features, and contextual state of EV $e_i$ at time $t$, and let $\mathbf{x}_{i,t}^{CP_j}$ denote the feature vector of candidate charging node $CP_j$ (e.g., distance, charging speed, popularity). The objective is to learn a parameterized scoring function $f_{\theta}(\cdot)$ that maps these features to a real-valued relevance score:

\begin{equation}
s_{i,j,t} = f_{\theta}\!\left(\mathbf{x}_{i,t}^{\text{context}},\, \mathbf{x}_{i,t}^{CP_j}\right),
\end{equation}

where $s_{i,j,t} \in \mathbb{R}$ denotes a real-valued relevance score that reflects the relative suitability of candidate charging node $CP_j$ for EV $e_i$ at decision time $t$. Higher values of $s_{i,j,t}$ indicate a stronger preference for selecting $CP_j$ relative to other candidate nodes within the same decision event.

Accordingly, the recommended charging nodes for EV $e_i$ at time $t$ are obtained by sorting candidates in descending order of their predicted scores:
\begin{equation}
\mathcal{CP}^{*}_{i,t}
=
\operatorname{argsort}_{\downarrow}
\left\{
s_{i,j,t}
\;:\;
CP_j \in \mathcal{CP}^{e_i}_t
\right\}.
\end{equation}

Model parameters $\theta$ are learned by minimizing a ranking loss 
$\mathcal{L}_{\mathrm{rank}}$, which measures the discrepancy between the predicted ordering of candidate charging nodes and the reference ordering defined by graded relevance labels $y_{i,j,t}$ within each EV decision event.

\begin{equation}
\min_{\theta}\; 
\mathcal{L}_{\mathrm{rank}}\!\left( \{ s_{i,j,t}(\theta) \}, \{ y_{i,j,t} \} \right).
\end{equation}

For a given decision event $i$ at time $t$, candidate charging nodes 
$j \in \mathcal{J}_i$ are correctly ordered when nodes with higher graded relevance labels receive higher predicted relevance scores, i.e.,
\[
y_{i,j,t} > y_{i,k,t}
\;\Rightarrow\;
s_{i,j,t} > s_{i,k,t},
\]
where $s_{i,j,t}$ denotes the predicted relevance score assigned to 
candidate node $j$. Hence, the graded labels $y_{i,j,t}$ defined a preference structure over candidate nodes, where larger values indicate stronger suitability for EV--EV energy trading under the corresponding mobility and energy context.

This formulation enables the model to learn an uncertainty-aware, event-specific ranking policy that generalizes beyond deterministic heuristics and supports robust charging node recommendation in EV--EV energy trading.

\section{Research methodology}
\label{rm}

\subsection{Overall system pipeline}

The process begins with the collection of multidimensional datasets, including taxi mobility traces, charging station specifications, and temporal popularity metrics of charging locations. These datasets are acquired and unified through a combination of API-based access and a custom integration algorithm. The aggregated data is then processed through cleaning, normalization, feature extraction, and feature selection stages. The selected features serve as input to the learning-to-rank model, which captures spatiotemporal dependencies in driver journey and trading dynamics (provider or consumer). To enhance model generalization and reduce overfitting, we apply hyperparameter tuning using manual search, along with regularization techniques. Then the trained model predicts the most probable charging locations, thereby supporting improved decision-making and operational efficiency in EV--EV energy trading.


\subsection{Data acquisition and preliminary analysis}
Next charging nodes recommendation for EV–EV energy trading requires datasets that jointly model EV journey and trading-related attributes, including dynamic trading roles (provider or consumer), energy transaction volumes, battery energy states, and the set of candidate charging locations encountered throughout an EV’s journey.

\begin{figure}
\centering
\small 
\includegraphics[width=0.55\textwidth]{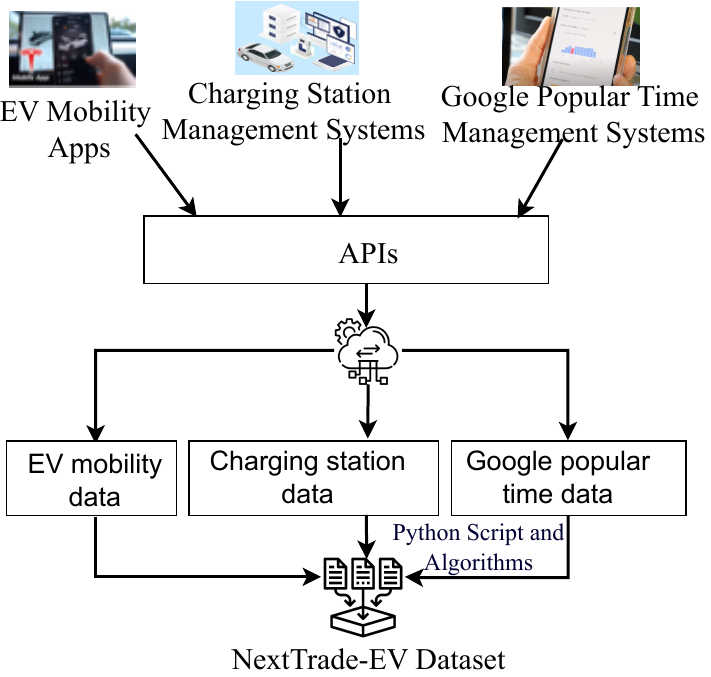}
\centering
\caption{Overall process for the EV--EV trading datasets generation.}
\label{fig:coordination-process}
\end{figure}


Our final dataset (NextTrade-EV) integrates four real-world empirical data sources: (i) Chicago taxi mobility traces, (ii) top-selling EV model from the USA and Australia, (iii) Chicago charging station infrastructure data, and (iv) Google popular times at charging locations. Building on these empirical observations, we further derive and augment physically grounded attributes, including EV energy levels estimated from travel distances and per-kilometer energy consumption rates. The complete set of dataset attributes used to train the ranking model is presented in Table \ref{tab:data-jorney-mobility}.

\begin{table*}
\scriptsize
\renewcommand{\arraystretch}{1.5} 
 \caption{Summary of Dataset Attributes for EV--EV energy trading}
 \begin{tabularx}{\textwidth}{p{22mm}p{15mm}p{15mm}p{95mm}}

 \hline Field & Type & Static/dynamic & Description\\ 

 \hline

EV journey ID & String & D & Unique identifier for each journey \\
 \hline

EV ID & String & S & Unique identifier for each EV \\
 \hline

EV Model name & String & S & An integer from 1 to 9 representing the EV model type allows easy vehicle category identification \\
 \hline

Battery capacity & int64 & S & Represents the maximum energy storage capacity of the EV’s battery, measured in watt-hours (Wh), indicating the total energy the battery can store. \\
 \hline

Battery range & int64 & D & Represent the maximum distance the EV can travel on a full charge, measured in kilometers. This value is reduced after every fare (trip) based on the km covered in the journey. \\
 \hline

Community area ID & int64 & S & An integer between 1 and 77, where each ID represents a designated community area in Chicago. Each community area is a geographic unit. \\
 \hline

Journey start and end time & Timestamp & D & Represent the start and end timestamps of each journey, used to track trip duration and temporal patterns. \\
 \hline

Time at locations  & Timestamp & D &  Represents specific timestamps when an EV arrives at or departs from key locations, including nearby charging stations, pickup/drop-off points, and intermediate travel stops. \\
 \hline

EV journey locations  & String & D &  Represents spatio-temporal journey points, including source, destination, and intermediate locations where EVs may stop, detour, trade, idle, or engage in travel interactions (trading, driving). \\
 \hline

Traveled distance  & int64 & D &  Represents the distance covered by the EV during a journey, measured in kilometers. \\
 \hline

Battery energy level  & int64 & D &  This feature represents the state of charge (SOC) of the EV's battery at various points in a specific time during the journey, measured in watt-hours (Wh). This is the most salient feature for understanding energy consumption, their role as a provider and consumer, estimating the expected energy exchange for trading, and analyzing the driver trading behavior. \\
 \hline

Energy consumption per journey  & int64 & D &  The total energy utilized by an EV for completing a single journey. This field is critical for understanding energy needs and planning trading opportunities. \\
 \hline

Candidate charging stations for trading & int64 & D &  A set of nearby charging stations identified as potential locations for energy trading. These stations are selected based on their proximity to the EV's journey path and are integrated into the driver's travel itinerary. 

\\ \hline

Distance to charging stations & int64 & D &  The spatial distance from an EV's current location to each candidate charging station. This metric impacts decisions regarding the feasibility and efficiency of visiting a station for energy trading. \\
 \hline

Charging speed of charging stations & int64 & D &  The charging rate related to candidate charging stations at which a charging station can transfer energy to or from an EV, influencing the duration of trading sessions. \\
 \hline

Historical density of EVs at charging locations & int64 & D &  Analogous to the concept of \emph{Google popular times} at charging locations, this metric reflects the historical availability of EVs at specific charging locations over time for energy trading. The purpose of \emph{Google popular times} is to provide users with insights about how busy a place is at different times of the day and week. This spatial context information uses aggregated and anonymized location data from users to estimate crowd levels at charging locations. In terms of energy implications, this information helps drivers make informed decisions about whether visiting a specific charging location is worthwhile for their needs. At charging nodes, \emph{popularity} 
reflects the relative level of human and vehicular activity around a station at a given time, and is used as a proxy for the likelihood of encountering potential trading partners.
 
\\

 \hline

Trading role & String & D & The dynamic role of an EV in energy trading, categorized as either a provider or a consumer. This attribute is crucial, as it enables the identification of trading dynamics (provider or consumer) at charging locations and helps optimize energy supply and demand management. \\
 \hline

Trading Volume & int64 & D & The amount of expected energy by providers and consumers in trading interaction, expressed in watt-hours (Wh).\\
 \hline

\end{tabularx}
 \label{tab:data-jorney-mobility}
 \end{table*}


To enable the use of large-scale taxi mobility data as a proxy for future EV–EV trading scenarios, we make two key assumptions. First, we assume that the majority of taxis will transition to EVs in the future. Second, we assume that all EVs employ a standardized wireless charging mechanism that is fully compatible with wireless charging pads. Under these assumptions, non-EV taxi trajectories can be treated as potential EV journeys, which simplifies dataset construction and feature derivation while remaining aligned with anticipated trends in electric mobility and charging infrastructure.

Our \emph{Chicago taxi mobility dataset} encompasses 77 community areas within the city of Chicago, each representing a distinct geographic region. It contains taxi journey records from 4,551 individual vehicles collected over a 12-month period, comprising more than 17 million journeys. Following the approach of Marlin \textit{et al.} \citep{C57}, we construct a synthetic EV mobility dataset from empirical taxi trajectories to preserve realistic spatio-temporal travel patterns while enabling the analysis of EV energy trading.

In this dataset, some entries in the \emph{trip distance} attribute were missing. To address this issue, we applied a two-stage imputation strategy. First, we used the Haversine formula \citep{winarno2017location} to estimate travel distances between pickup and drop-off locations based on their geographic coordinates. Second, for journeys where coordinate-based distance estimation was unavailable, we computed an average fare-per-kilometer rate from completed trips and used this rate to estimate missing distances based on the corresponding fare values. This preprocessing step ensured consistency and completeness of distance-related features required for subsequent mobility and charging analysis.

\begin{figure}
        \centering
        \small 

        \includegraphics[width=0.8\textwidth]{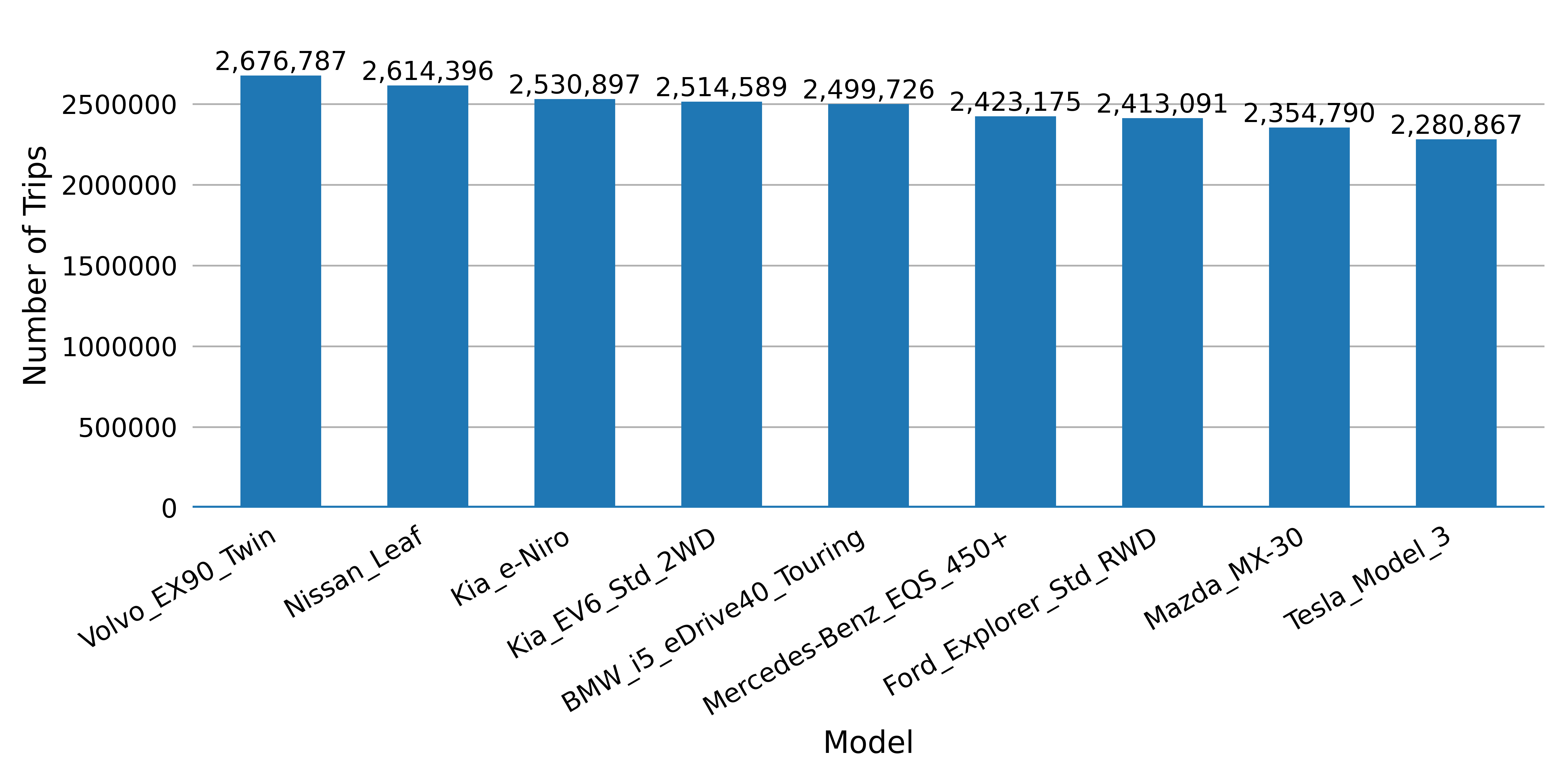}
        
        \centering
        \caption{Distribution of trips per EV model shown as a bar plot, ordered from most to least frequent model.}
        \label{fig:ev-model-dist}
        \end{figure}
\begin{figure}
        \centering
        \small 

        \includegraphics[width=0.8\textwidth]{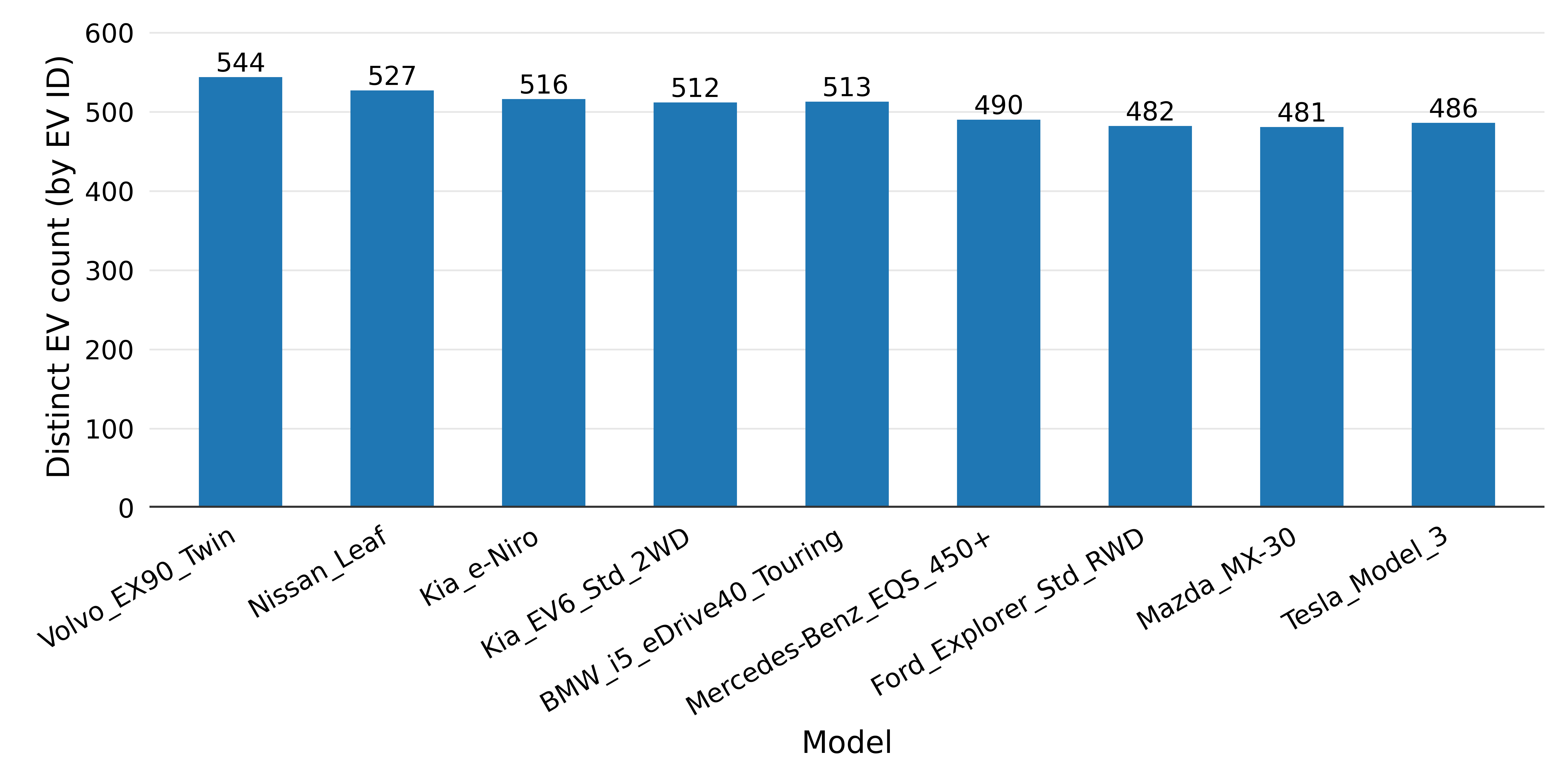}
        
        \centering
        \caption{Distribution of unique EVs per model (distinct EV IDs), ordered by model frequency.}
        \label{fig:ev-model-dist-unique}
        \end{figure}

As a second empirical data source, we incorporate \emph{EV model datasets} containing model names, battery capacities, and total driving range attributes, which are integrated with the previously described taxi mobility data to map each taxi journey to EV-specific characteristics. Each taxi ID is first mapped to a unique EV ID, ensuring that all journeys associated with a given taxi correspond to the same EV. Subsequently, EV IDs are randomly assigned to one of nine EV models using a uniform distribution over model types and their corresponding attributes. This random assignment assumes that each EV model is represented by a sufficiently large training sample, as illustrated in Figures \ref{fig:ev-model-dist} and \ref{fig:ev-model-dist-unique}. As a result, the assignment process remains unbiased.



\begin{figure*}
\centering
\subfloat[]{\includegraphics[width=0.50\textwidth]{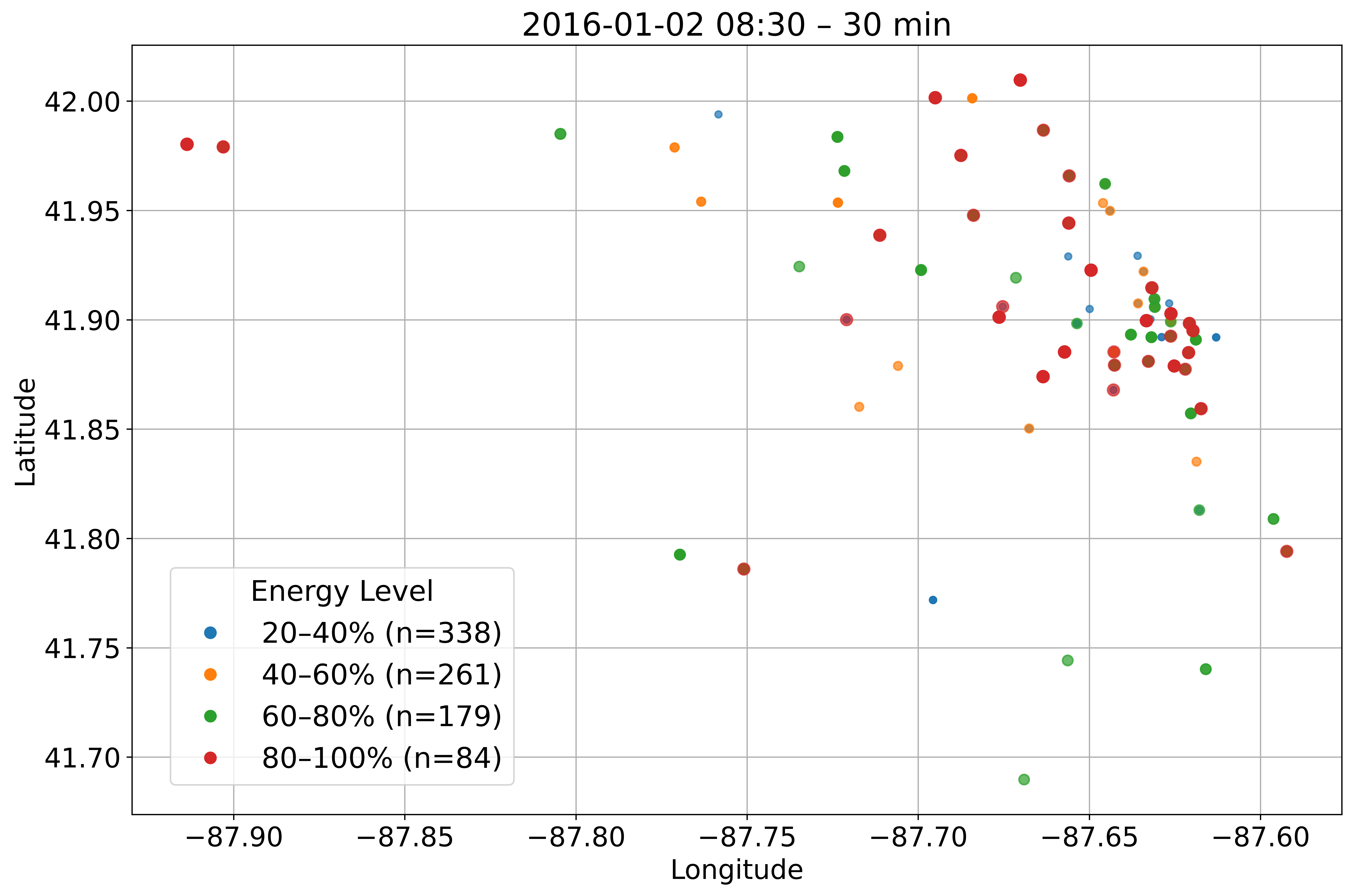}}\hfill
\subfloat[]{\includegraphics[width=0.50\textwidth]{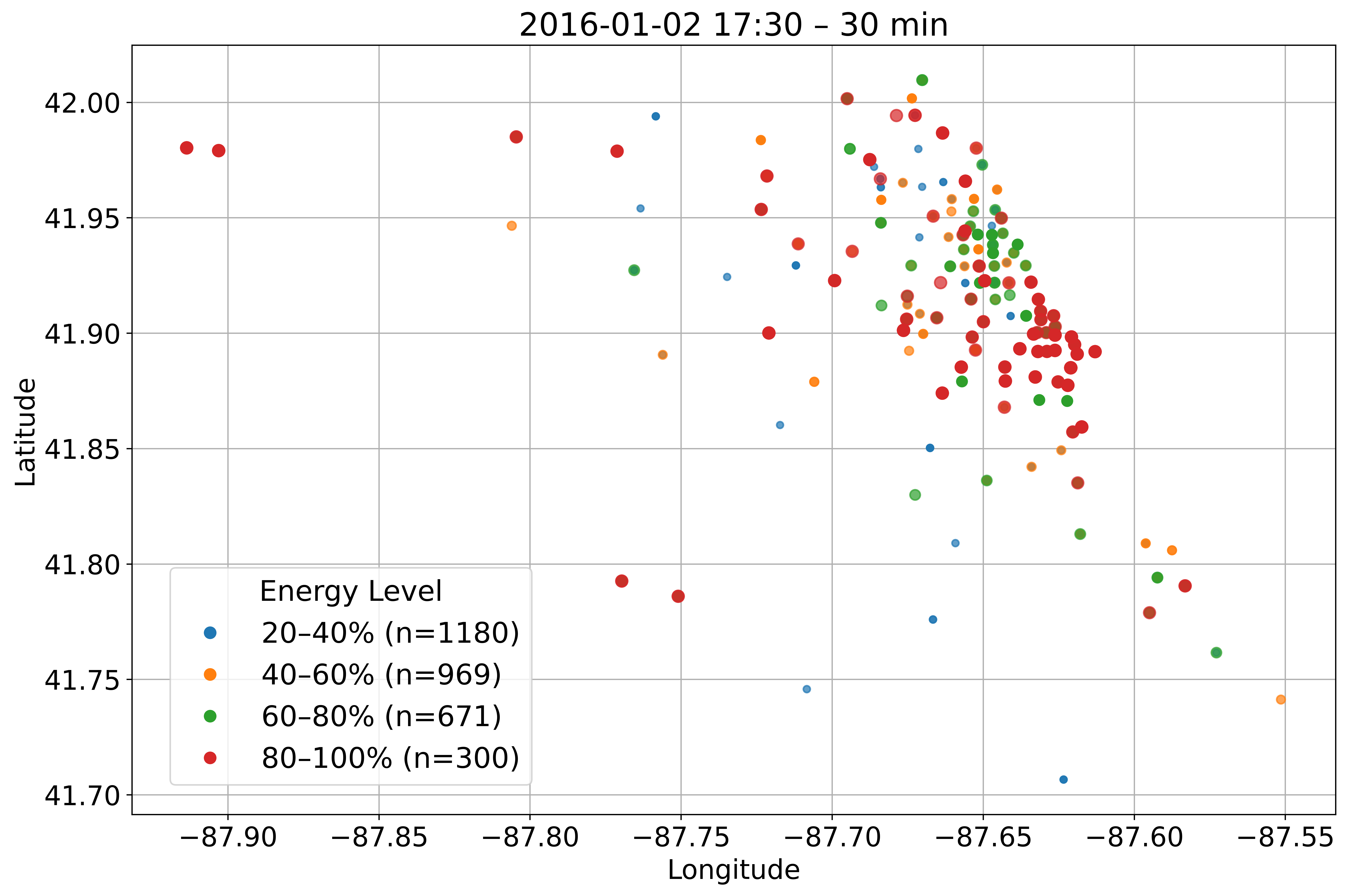}}\\[0.6em]
\subfloat[]{\includegraphics[width=0.50\textwidth]{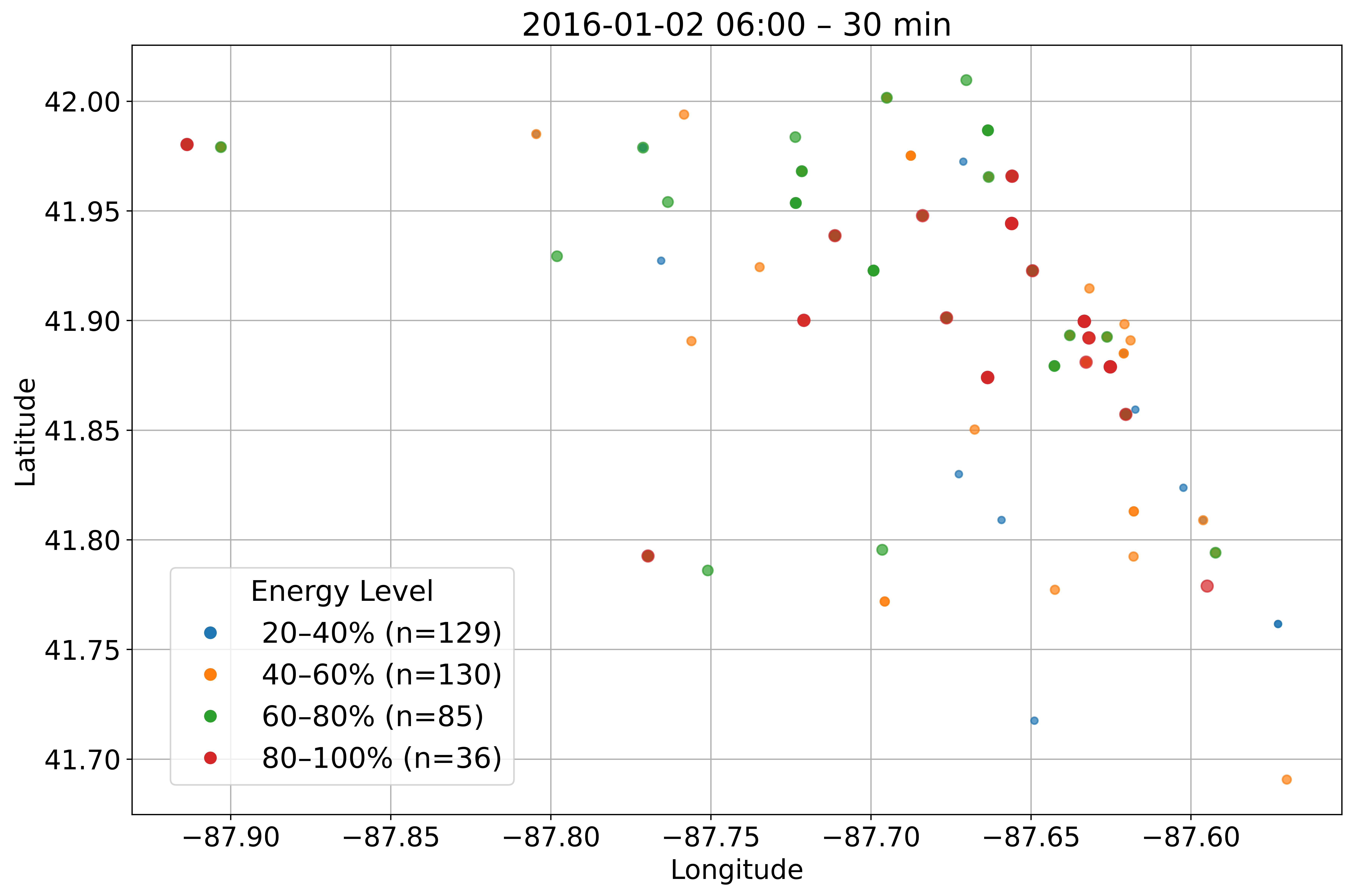}}\hfill
\subfloat[]{\includegraphics[width=0.50\textwidth]{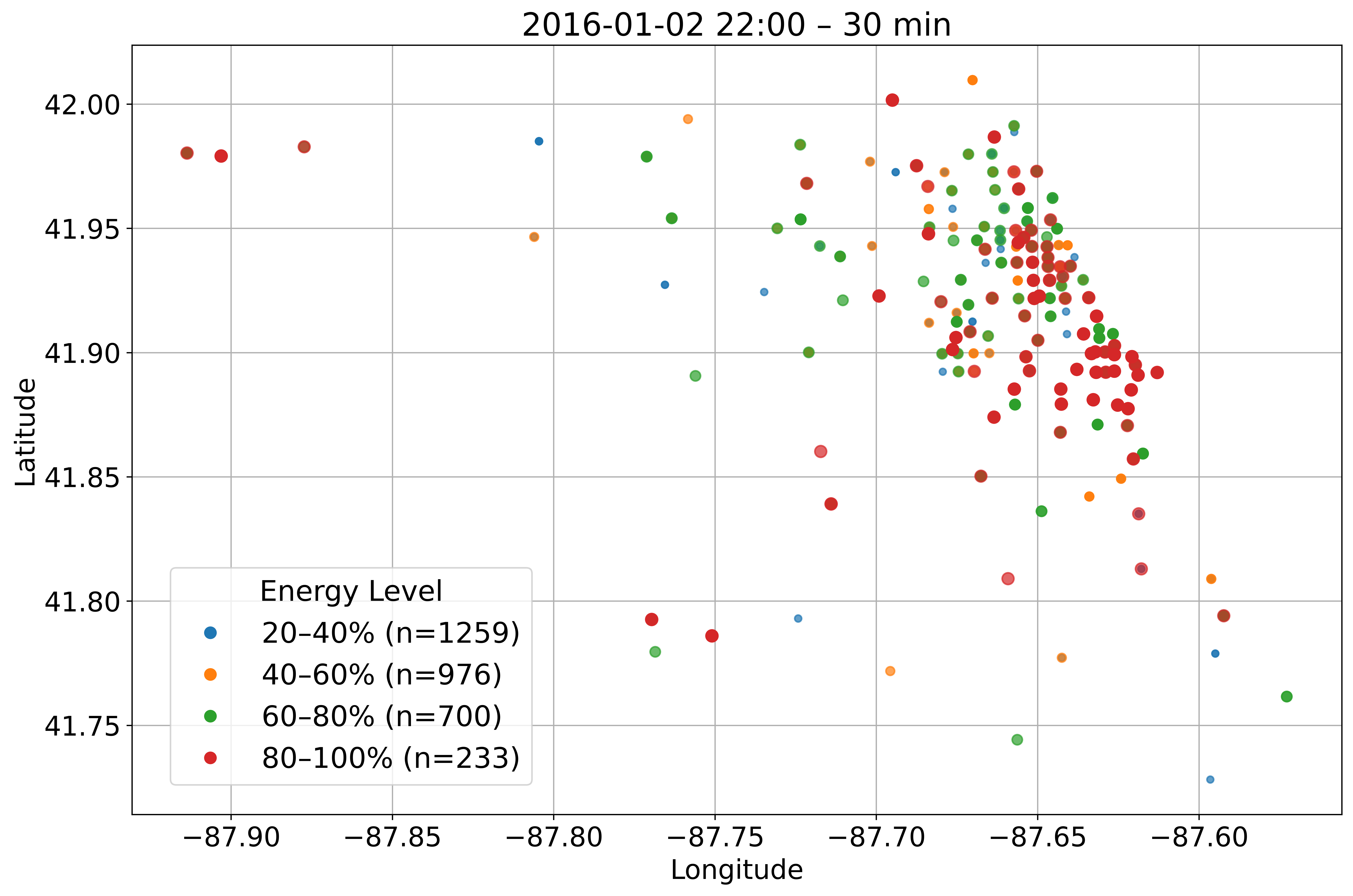}}
\caption{Spatial distribution of EV locations by energy level (20–40\%, 40–60\%, 60–80\%, 80–100\%) for 30-minute windows, high peak (08:30, 17:30) and off-peak (06:00, 22:00) periods; legends report per energy level counts.}
\label{fig:ev-soc}
\end{figure*}

The \emph{energy levels} at the source and destination are dynamically generated based on the EV's initial battery state and the energy consumed during the journey. The initial battery energy level for each EV was determined using a \emph{uniform distribution} between 20\% and 100\% of its maximum battery capacity. This method ensures that each EV's starting energy level is randomly distributed across the specified range, reflecting realistic variations in the state of charge. During the simulation, energy consumption for each journey is calculated based on the distance traveled and the standard EV model's energy consumption rate. After completing a journey, the destination energy level is determined by subtracting the energy consumed from the initial energy level. Additionally, when an EV’s energy level drops below a predefined threshold (20\% of its battery capacity), a recharging event is synthetically generated following the charge-feature construction procedure of Marlin \textit{et al.} \citep{C57}. Specifically, the battery charge level is decremented according to the distance traveled and an energy-per-kilometer consumption rate, and once the 20\% threshold is reached, the battery is assumed to be recharged to full capacity before the vehicle continues its subsequent journey segment. This recharging process is used only to maintain realistic energy feasibility in the synthetic EV mobility data and does not represent a modeled charging location or charging decision. Accordingly, the location of this synthetic recharging event is not explicitly modeled and is independent of the charging node recommendation problem studied in this study. Figure \ref{fig:ev-soc} is included to illustrate the underlying data distribution, showing how the number of EVs at different energy levels varies across time windows, which is relevant for understanding energy-dependent charging behavior and informing energy-aware charging node recommendation.


        

Integration of \emph{charging station} data, including the \emph{distance} from each EV’s location, is a crucial step in identifying suitable charging options. We collected charging station information for the same geographic region as the taxi mobility data using multiple APIs. The primary data sources include \emph{Google Maps} (\url{https://www.google.com/maps}
), \emph{EV Charger Finder} (\url{https://ev-charger-finder.p.rapidapi.com/}
), and the \emph{City of Chicago Open Data Portal} (\url{https://data.cityofchicago.org/}
). 
To identify nearby candidate stations efficiently for each driver’s journey, we developed an adaptive search procedure presented in Algorithm \ref{alg:optimized_nearby_station}. The algorithm dynamically expands the search radius--from 1 km up to 10 km--until at least three charging stations are found for each location along the EV’s journey route. The proximity between each EV location and station is calculated using the Haversine distance formula \citep{sinnott1984virtues, winarno2017location}. The distance calculation is shown in Eq.~\eqref{haversine}. Let $(\phi_1, \lambda_1)$ and $(\phi_2, \lambda_2)$ denote the latitude and longitude (in radians) of an EV location and a charging point, respectively, and let $R$ be the Earth’s mean radius (approximately $6371$ km). The Haversine distance $d$ between the two points is computed as:

\begin{equation}
\label{haversine}
\begin{aligned}
\Delta \phi &= \phi_2 - \phi_1, \qquad
\Delta \lambda = \lambda_2 - \lambda_1, \\
a &= \sin^2\!\left(\frac{\Delta \phi}{2}\right)
     + \cos(\phi_1)\cos(\phi_2)\sin^2\!\left(\frac{\Delta \lambda}{2}\right), \\
d &= 2R \arctan2\!\left(\sqrt{a}, \sqrt{1-a}\right).
\end{aligned}
\end{equation}

Here, $d$ represents the great-circle distance between the EV and the charging point along the Earth’s surface. This adaptive and data-driven approach ensures comprehensive coverage across dense and sparse urban regions, ultimately producing an enriched dataset that links every journey with its set of candidate charging locations and corresponding distances. After integrating these candidate stations into the EV journey dataset, we further enriched the data with key station attributes, such as the \emph{popularity index of charging locations over time} and the \emph{charging speed} of each station.

Algorithm \ref{alg:energy_volume_role} assigns a trading role (provider or consumer) and a corresponding energy quantity to each EV journey based on its energy condition. Building on the previously defined battery capacity $B_c^{e_i}$ and available surplus energy $E_{e_i}^{\mathrm{sur}}(t)$ (computed according to Eq. \ref{eq:avail_prov}), we derive a scale-invariant measure of trading potential by normalizing the surplus energy with respect to the EV's own battery capacity: 


\begin{equation}
r = \min\left\{1,\; \frac{E_{e_i}^{\mathrm{sur}}(t))}{ B_c^{e_i}}\right\}.
\end{equation}

This ratio $r \in [0,1]$ represents the portion of the EV’s battery capacity that can be safely traded without violating the minimum operational state-of-charge (SoC) required for the EV to complete its own trip. Importantly, this normalization is self-referential, meaning that the ratio is computed relative to the same EV’s capacity rather than another vehicle’s capacity. Consequently, $r$ captures relative surplus availability independent of absolute battery size differences 
across EV models.

The trading role $R_j \in \{\mathrm{P}, \mathrm{C}\}$ is then assigned according to the following piecewise probabilistic rule:
\begin{equation}
R_j =
\begin{cases}
\mathrm{C}, & 0 \le r \le 0.30, \\[4pt]

\begin{aligned}
\mathrm{P} &\text{ with probability } p_{\text{mid}},\\
\mathrm{C} &\text{ otherwise},
\end{aligned}
& 0.30 < r \le 0.70, \\[8pt]

\begin{aligned}
\mathrm{P} &\text{ with probability } p_{\text{high}},\\
\mathrm{C} &\text{ otherwise},
\end{aligned}
& 0.70 < r \le 0.90, \\[8pt]

\mathrm{P}, & r > 0.90.
\end{cases}
\end{equation}
For intermediate surplus ranges, role assignment is probabilistic.
Specifically, a binary random variable $X$ is generated such that:
\[
\mathbb{P}(X = 1) = p, 
\qquad 
\mathbb{P}(X = 0) = 1 - p,
\]
Where $X = 1$ corresponds to assigning the role $\mathrm{P}$
(provider) and $X = 0$ corresponds to assigning the role $\mathrm{C}$
(consumer). The probability parameter $p$ takes the value
$p_{\text{mid}}$ or $p_{\text{high}}$ depending on the surplus band.
The probabilistic assignment in intermediate regions models EV driver behavioral uncertainty: although sufficient surplus exists, willingness to trade may depend on contextual or preference-related factors not explicitly observed in the dataset.

After determining the journey-level role $R_j$, the corresponding
energy quantity $Q_j$ is computed in a role-dependent manner.
Specifically,
\begin{equation}
Q_j =
\begin{cases}
\max\{0,\, E_{e_i}^{\mathrm{sur}}(t)\}, 
& \text{if } R_j = \mathrm{P}, \\[6pt]
E_{e_i}^{\mathrm{cons}}(t), 
& \text{if } R_j = \mathrm{C}.
\end{cases}
\end{equation}

For provider EVs, trades below the minimum threshold
$E_{\min}^{\mathrm{trade}}$ are suppressed to avoid negligible
energy exchanges, i.e.,
\begin{equation}
Q_j = 0 \quad \text{if } R_j=\mathrm{P}
\text{ and } Q_j < E_{\min}^{\mathrm{trade}}.
\end{equation}

Finally, the trading role $R_j$ and quantity $Q_j$ are replicated across all $L$ timestamps of the journey to produce time-aligned lists used in the EV--EV trading simulation.

\begin{algorithm}
\scriptsize
\caption{Finding Adaptive Nearby Candidate Charging Locations for Each EV Journey}
\label{alg:optimized_nearby_station}
\KwIn{$df\_journey$: DataFrame of EV journeys with locations, $df\_station$: DataFrame of charging stations with latitude, longitude, and station IDs}
\KwOut{Updated $df\_journey$: DataFrame with nearby charging stations and distances}

\BlankLine
Load journey data $df\_journey$ and charging station data $df\_station$;
\BlankLine
Initialize empty lists: $nearby\_stations\_list, distances\_list$;

\ForEach{$ev\_route$ in $df\_journey$}{
    Convert route string to a list of tuples $(latitude, longitude)$;
    
    Initialize empty lists: $route\_nearby\_stations$, $route\_distances$;
    
    \ForEach{$ev\_location$ in $ev\_route$}{
        $search\_radius \gets 1, \quad min\_stations = 3$;
        
        \While{$search\_radius \leq 10$}{
            Initialize empty lists: $nearby\_stations, distances$;
            
            \ForEach{station in $df\_station$}{   
                Compute \textbf{Haversine distance} $d$ between $ev\_location$ and $station['station\_location']$;   
                
                \If{$d \leq search\_radius$}{  
                    Append $(station\_id, d)$ to respective lists;
                }
            }
            
            Remove duplicates and ensure at least 3 stations are found;
            
            \If{$|nearby\_stations| \geq min\_station\_count$}{
                \textbf{break}\;
            }
            \Else{
                Increase search radius: $search\_radius \gets search\_radius + 1$\;
            }
        }

        Append $(nearby\_stations, distances)$ to $route\_nearby\_stations$, $route\_distances$ lists;        
                
    }

    Append ($route\_nearby\_stations$, $route\_distances$) to $nearby\_stations\_list, distances\_list$;
}
Assign $nearby\_stations\_list$ to $df\_journey$['$candidate\_charging\_stations$'] and $distances\_list$ to $df\_journey$['$distance\_to\_cs$'];

Save $df\_journey$ as CSV\;
\Return $df\_journey$;
\end{algorithm}

\begin{algorithm}
\scriptsize
\caption{Per-Journey Role Assignment (Provider and Consumer) and Energy Volume}
\label{alg:energy_volume_role}
\KwIn{Dataset $\mathcal{D}$ with columns: \texttt{battery\_energy\_level} (list, Wh), 
\texttt{battery\_capacity} (Wh), 
\texttt{time\_at\_locations} (list)}
\KwIn{$E_{\min}^{\mathrm{trade}}{=}10000$ (Wh), 
$p_{\text{mid}}{=}0.50$, 
$p_{\text{high}}{=}0.80$}
\KwOut{Dataset $\mathcal{D}'$ with \texttt{roles} (list) and \texttt{quantity\_of\_energy} (list)}

Parse list-type columns from strings to lists\;
Optionally set a NumPy seed for reproducibility\;
Initialize $\mathcal{D}' \gets \emptyset$\;

\ForEach{journey $j$ in $\mathcal{D}$}{
  $B \gets \texttt{battery\_capacity}[j]$\;
  $\mathcal{T} \gets \texttt{time\_at\_locations}[j]$\;
  $L \gets |\mathcal{T}|$\;

  \textbf{Step 1: Compute self-normalized surplus ratio}\\
  $E_{e_i}^{\mathrm{sur}}(t) \gets$ available surplus energy using Eq.~(\ref{eq:avail_prov})\;
  $r \gets \min\{1,\; E_{e_i}^{\mathrm{sur}}(t)/ B_c^{e_i}\}$\;

  \textbf{Step 2: Role assignment}\\
  \uIf{$0 \le r \le 0.30$}{
    $R_j \gets \text{`C'}$\;
  }
  \uElseIf{$0.30 < r \le 0.70$}{
    $R_j \gets \text{binomial}(1, p_{\text{mid}}) ? \text{`P'} : \text{`C'}$\;
  }
  \uElseIf{$0.70 < r \le 0.90$}{
    $R_j \gets \text{binomial}(1, p_{\text{high}}) ? \text{`P'} : \text{`C'}$\;
  }
  \Else{
    $R_j \gets \text{`P'}$\;
  }

  \textbf{Step 3: Quantity assignment}\\
  $E^{\mathrm{cons}}_{j}(t) \gets$ consumer requirement using Eq.~(\ref{cons})\;

  \uIf{$R_j = \text{`P'}$}{
    $Q_j \gets \max\{0,\; E^{\mathrm{sur}}_{j}(t)\}$\;
    \If{$Q_j < E_{\min}^{\mathrm{trade}}$}{
      $Q_j \gets 0$\;
    }
  }
  \Else{
    $Q_j \gets E^{\mathrm{cons}}_{j}(t)$\;
  }

  \textbf{Step 4: Broadcast to timestamps}\\
  \texttt{roles}$[j] \gets [R_j] \times L$\;
  \texttt{quantity\_of\_energy}$[j] \gets [\text{round}(Q_j,2)] \times L$\;

  Add updated journey $j$ to $\mathcal{D}'$\;
}
Save $\mathcal{D}'$ to CSV\;
\end{algorithm}

\subsection{Label selection}
 
\begin{figure*}
\centering
\small 
\includegraphics[width=\textwidth]{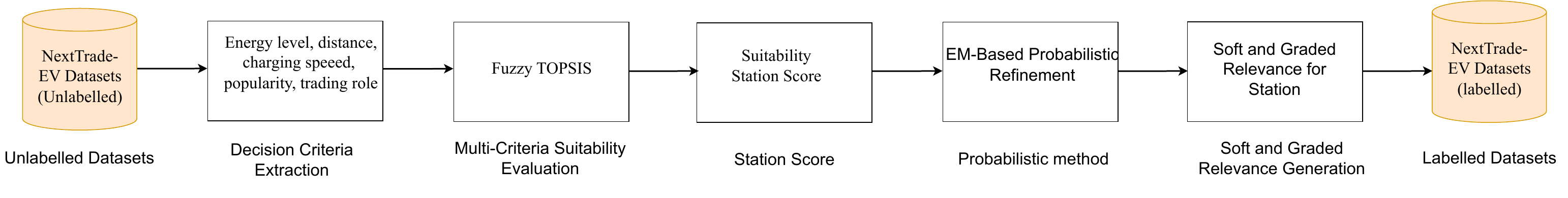}
\caption{Label generation pipeline.
Unlabeled EV decision events from the NextTrade-EV dataset are first evaluated using fuzzy TOPSIS to compute multi-criteria station suitability scores. These scores are subsequently refined through EM-based probabilistic modeling to capture latent contextual uncertainty. The resulting soft relevance scores are discretized into graded relevance labels, yielding a weakly labeled dataset suitable for learning-to-rank without requiring observed station selection data.}
\label{fig:coordination}
\end{figure*}

In the prediction model, we consider candidate charging stations as the possible charging locations for EV--EV energy trading. For the purpose of the training, the model requires both positive and negative samples as labels (relevance grades) \citep{sun2024multi, jeunen2023gradient, covington2016deep, oard1998implicit}. However, explicit EV–EV trading selections for charging nodes are not directly observable in historical mobility data. Relevance labels are constructed probabilistically to reflect the relative plausibility of charging nodes being considered for energy trading under the observed mobility and energy context. Figure~\ref{fig:coordination} presents the overall label generation pipeline. We apply the following steps for label construction. 


\subsubsection{Fuzzy-weighted TOPSIS for station suitability estimation}
\label{sec:step2_fuzzy_topsis}

At each decision event $e$, corresponding to an EVs at time $t$, a finite set of candidate charging stations $\mathcal{J}_e$ is available after geographic filtering. In EV--EV energy trading, station selection is conditioned on the EV’s trading role: \emph{consumer} EVs prioritize reliable energy acquisition, while \emph{supplier} EVs aim to efficiently transfer surplus energy. These asymmetric objectives, together with uncertainty in operational conditions, motivate a multi-criteria decision formulation under imprecision. Accordingly, we model station evaluation as a fuzzy multi-criteria decision-making (MCDM) problem and employ a fuzzy-weighted TOPSIS (Technique for Order Preference by Similarity to Ideal Solution) approach to compute a suitability score for each candidate station. TOPSIS ranks alternatives by evaluating their geometric proximity to an ideal solution and their distance from the worst-case solution in a multi-criteria space. In our formulation, fuzziness is applied to the \emph{importance weights of the decision criteria}, rather than to the numerical values of WCP attributes. WCP attributes correspond to physically measurable and operational quantities such as geographic distance (km), charging or discharging rate (kW), and availability (percentage) are treated as deterministic. Fuzziness is instead used to model uncertainty in the relative importance of these criteria, which may vary across EV roles, energy urgency, and changing energy level.

At each decision event $e$, corresponding to an EV at time $t$, a finite set of candidate charging points (CPs), denoted by $\mathcal{J}_e$, is available after geographic filtering. In EV--EV energy trading, CP selection is conditioned on the EV’s trading role: \emph{consumer} EVs prioritize reliable energy acquisition, while \emph{supplier} EVs aim to efficiently transfer surplus energy. These asymmetric objectives, together with uncertainty in operational conditions and driver preferences, motivate a multi-criteria decision formulation under imprecision.

Accordingly, we model CP evaluation as a fuzzy multi-criteria decision-making (MCDM) problem and employ a fuzzy-weighted TOPSIS (Technique for Order Preference by Similarity to Ideal Solution) approach to compute a suitability score for each candidate CP. TOPSIS ranks alternatives by evaluating their geometric proximity to an ideal solution and their distance from a worst-case solution in a multi-criteria space. In our formulation, fuzziness is applied exclusively to the \emph{importance weights of decision criteria}, rather than to the numerical values of CP attributes (e.g., distance, charging rate, or availability). This design choice captures uncertainty in decision priorities while maintaining deterministic and comparable attribute values, thereby avoiding distortion of physical quantities and ensuring numerical stability in the TOPSIS distance calculations.

\textbf{Evaluation criteria:} Each candidate station $j \in \mathcal{J}_e$ is characterized using the following criteria:
\begin{itemize}
    \item \emph{Distance} $d_{e,j}$ (cost criterion), representing the travel effort required to reach the station at time $t$.
    \item \emph{Charging speed / transfer capacity} $s_{e,j}$ (benefit criterion), capturing the potential transaction throughput.
    \item \emph{Popularity / availability} $a_{e,j}$ (benefit criterion), indicating the likelihood of successful access or matching at time $t$.
\end{itemize}

These criteria jointly capture spatial, temporal, and market-level factors relevant to station suitability in EV--EV energy trading.

\textbf{Event-wise normalization and cost--benefit conversion:} Because each decision event has its own candidate set, all criteria are normalized \emph{within each event}. For benefit-type criteria $x \in \{s,a\}$, min--max normalization is applied:

\begin{equation}
x^{\mathrm{norm}}_{e,j}
=
\frac{x_{e,j}-\min_{k\in\mathcal{J}_e} x_{e,k}}
{\max_{k\in\mathcal{J}_e} x_{e,k}-\min_{k\in\mathcal{J}_e} x_{e,k}+\epsilon}.
\end{equation}

Distance is a cost criterion and is converted into a benefit-type measure via min--max normalization followed by inversion:

\begin{equation}
d^{\mathrm{norm}}_{e,j}
=
1-
\frac{d_{e,j}-\min_{k\in\mathcal{J}_e} d_{e,k}}
{\max_{k\in\mathcal{J}_e} d_{e,k}-\min_{k\in\mathcal{J}_e} d_{e,k}+\epsilon}.
\end{equation}

Here, $\epsilon$ is a small positive constant introduced to avoid degenerate cases when all candidates exhibit identical values for a criterion, in which case the criterion becomes non-discriminative within the event. If $\max_{k\in\mathcal{J}_e} x_{e,k}=\min_{k\in\mathcal{J}_e} x_{e,k}$ for any criterion $x\in\{d,s,a\}$, we set the corresponding normalized values to $x^{\mathrm{norm}}_{e,j}=0.5$ for all $j\in\mathcal{J}_e$, rendering that criterion non-discriminative for event $e$. After transformation, all normalized criteria lie in $[0,1]$, with higher values indicating better performance. 

\textbf{Trading transaction pressure:} To encode role-dependent urgency in EV--EV trading, we define a transaction pressure variable $p_e \in [0,1]$ based on the EV's state-of-charge (SOC):
\begin{equation}
\mathrm{SOC}_e = \mathrm{clip}\!\left(\frac{E_e}{Cap_e},0,1\right),
\end{equation}
\begin{equation}
p_e =
\begin{cases}
1-\mathrm{SOC}_e, & \text{consumer EV},\\
\mathrm{SOC}_e, & \text{supplier EV}
\end{cases}
\end{equation}

This formulation captures minimal, domain-consistent constraints: consumers exhibit increasing urgency as SOC decreases, suppliers exhibit increasing propensity to trade as SOC increases. This pressure variable $p_e$ captures contextual urgency rather than estimated preference or observed behavior and is used only to modulate criterion importance.

\textbf{Fuzzy weighting of criteria:} Criterion importance under EV--EV trading is uncertain and varies with the operational context. We represent criterion importance using triangular fuzzy numbers (TFNs) corresponding to three ordinal levels:
\begin{align}
\text{High} &= (0.7,\,0.9,\,1.0), \\
\text{Medium} &= (0.3,\,0.5,\,0.7), \\
\text{Low} &= (0.1,\,0.3,\,0.5).
\end{align}
The overlap among the triangular fuzzy numbers is intentional and reflects gradual transitions between linguistic importance levels, which is a standard practice in fuzzy MCDM to model uncertainty and avoid brittle decision boundaries. These fuzzy representations encode bounded uncertainty while preserving monotonic dominance (High $>$ Medium $>$ Low).

Three fuzzy pressure regimes $g\in\{\text{low},\text{med},\text{high}\}$ are defined over the transaction pressure variable $p_e$ using standard triangular membership functions, which assign membership degrees $\mu_g(p_e)\in[0,1]$. Specifically, the low, medium, and high pressure regimes are defined by the triangular functions $(0,0,0.5)$, $(0.25,0.5,0.75)$, and $(0.5,1,1)$, respectively.
For each EV role $\rho(e)\in\{\text{consumer},\text{supplier}\}$, we specify defuzzified regime-level criterion importance profiles $\tilde{w}_{\rho,g,c}$ for $c\in\{d,s,a\}$ (distance, speed, availability), reflecting domain-consistent priorities under different pressure regimes. The triangular fuzzy number associated with linguistic importance levels is defuzzified by taking its centroid, resulting in a single relative numerical importance weight (a crisp regime-level weight).

The regime-level weights $\tilde{w}_{\rho,g,c}$ are normalized importance proportions derived from defuzzified linguistic priorities, constructed to preserve relative dominance and monotonic shifts across pressure regimes rather than computed through an analytic formula.

Event-specific criterion weights $w_{e,c}$ are obtained by membership-weighted interpolation of the regime-level weights and subsequent normalization:
\begin{equation}
w_{e,c}
=
\frac{\sum_{g}\mu_g(p_e)\,\tilde{w}_{\rho(e),g,c}}
{\sum_{c'\in\{d,s,a\}}\sum_{g}\mu_g(p_e)\,\tilde{w}_{\rho(e),g,c'}} ,
\qquad
\sum_{c} w_{e,c}=1.
\end{equation}

\textbf{TOPSIS scoring via geometric proximity:} Let $x^{\mathrm{norm}}_{e,j,c}$ denote the normalized value of criterion $c\in\{d,s,a\}$. The weighted normalized representation is given by:
\begin{equation}
v_{e,j,c} = w_{e,c}\, x^{\mathrm{norm}}_{e,j,c}.
\end{equation}

Since all criteria are benefit-type after conversion, the positive and negative ideal solutions are defined as:
\begin{equation}
v^{+}_{e,c} = \max_{j\in\mathcal{J}_e} v_{e,j,c}, \quad
v^{-}_{e,c} = \min_{j\in\mathcal{J}_e} v_{e,j,c}.
\end{equation}

The Euclidean distances to the ideal and anti-ideal solutions are:
\begin{align}
D^{+}_{e,j} &= \sqrt{\sum_c \left(v_{e,j,c}-v^{+}_{e,c}\right)^2}, \\
D^{-}_{e,j} &= \sqrt{\sum_c \left(v_{e,j,c}-v^{-}_{e,c}\right)^2}.
\end{align}

The station suitability score is computed using the TOPSIS closeness coefficient:
\begin{equation}
r_{e,j} = \frac{D^{-}_{e,j}}{D^{+}_{e,j}+D^{-}_{e,j}}, \quad r_{e,j}\in[0,1].
\end{equation}

A higher value of $r_{e,j}$ indicates that the station is closer to the ideal multi-criteria compromise and farther from the worst-case alternative.
These scores serve as a decision prior and are subsequently refined through probabilistic smoothing and supervised learning-to-rank in later stages.

\subsubsection{EM-based probabilistic smoothing of station suitability}
\label{sec:step3_em}

Let $e$ denote an EV decision event and $\mathcal{J}_e$ the corresponding set of candidate charging or trading stations. The fuzzy-weighted TOPSIS method calculates a deterministic suitability score $r_{e,j}\in[0,1]$ for each candidate $j\in\mathcal{J}_e$. 
While some decision events exhibit a clearly dominant station, many EV–EV energy trading scenarios involve dense infrastructure, competition among providers, or role-dependent urgency asymmetries between consumers and suppliers, leading to multiple stations with comparable suitability scores.
Moreover, the dataset does not provide observed station selection labels. To account for such ambiguity and improve robustness, we apply a probabilistic smoothing stage based on the Expectation--Maximization (EM) algorithm.

\textbf{Observed data and scope:}
The EM model operates exclusively on the suitability scores $\{r_{e,j}\}$ produced through fuzzy TOPSIS. It does not directly use the original dataset features, which are implicitly encoded in $r_{e,j}$. This separation avoids double-counting and preserves modularity between heuristic decision modeling and probabilistic refinement.

\textbf{Latent regime formulation:}
We assume that station suitability scores arise from a small number of unobserved but recurring contextual patterns, referred to as \emph{latent regimes}. These regimes do not represent driver types or behavioral classes; rather, they capture distinct EV--EV energy trading contexts as reflected in the distributional shape of suitability scores (e.g., sharply peaked distributions indicating clear dominance among candidates versus flatter distributions indicating high decision ambiguity).

Formally, the marginal distribution of suitability scores is modeled as a finite mixture:
\begin{equation}
p(r_{e,j}) = \sum_{k=1}^{K} \pi_k \, p(r_{e,j} \mid z_{e,j}=k),
\end{equation}
where $z_{e,j}$ is a latent regime indicator and $\pi_k$ denotes the mixture weight, with $\sum_{k=1}^{K}\pi_k = 1$. The latent regimes are shared across all decision events and capture recurring distributional patterns in station suitability scores rather than event-specific or driver-specific states.

\textbf{Likelihood model:}
Since $r_{e,j}\in[0,1]$, each latent regime is modeled using a Beta distribution:
\begin{equation}
p(r_{e,j}\mid z_{e,j}=k) = \mathrm{Beta}(r_{e,j}\mid \alpha_k,\beta_k),
\end{equation}
The Beta distribution is employed because station suitability scores are continuous and bounded in $[0,1]$, and its flexible shape allows representation of both decisive and ambiguous trading contexts.

\textbf{EM estimation:}
Given current parameter estimates $\{\pi_k,\alpha_k,\beta_k\}$, the E-step computes posterior responsibilities:
\begin{equation}
\gamma_{e,j,k}
=
\frac{
\pi_k \,\mathrm{Beta}(r_{e,j}\mid \alpha_k,\beta_k)
}{
\sum_{\ell=1}^{K}
\pi_\ell \,\mathrm{Beta}(r_{e,j}\mid \alpha_\ell,\beta_\ell)
}.
\end{equation}
In the M-step, mixture weights are updated as
\begin{equation}
\pi_k
=
\frac{1}{N}
\sum_{e}\sum_{j\in\mathcal{J}_e}
\gamma_{e,j,k},
\end{equation}
where $N=\sum_e |\mathcal{J}_e|$ is the total number of event--station pairs. The Beta parameters $(\alpha_k,\beta_k)$ are updated by maximizing the expected complete-data log-likelihood. The number of regimes $K$ is selected from a small range (e.g., $2\le K\le5$) to balance expressiveness and stability.

\textbf{Smoothed suitability estimation:}
Each latent regime $k$ is associated with an expected suitability level given by the mean of its Beta distribution:
\begin{equation}
\mu_k = \mathbb{E}[r \mid z=k] = \frac{\alpha_k}{\alpha_k+\beta_k}.
\end{equation}
For each event--station pair $(e,j)$, we compute a smoothed suitability score as the responsibility-weighted expectation:
\begin{equation}
\hat r_{e,j} = \sum_{k=1}^{K} \gamma_{e,j,k}\,\mu_k .
\end{equation}

\textbf{Event-level soft relevance:}
For each event $e$, the smoothed suitability scores are normalized across the candidate set to obtain a soft relevance distribution:
\begin{equation}
P_{e,j}
=
\frac{
\hat r_{e,j}
}{
\sum_{j'\in\mathcal{J}_e} \hat r_{e,j'}
},
\qquad
\sum_{j\in\mathcal{J}_e} P_{e,j} = 1.
\end{equation}
The values $P_{e,j}$ quantify relative plausibility under uncertainty and do not represent observed station choice probabilities.

EM is employed solely as a probabilistic smoothing mechanism. It does not model driver behavior or explicit market dynamics and does not rely on observed station selection labels.

\textbf{Graded relevance generation:}
\label{sec:step4_grades}
The event-level soft relevance scores $P_{e,j}$ produced in EM smoothing capture uncertainty in station suitability. However, most learning-to-rank algorithms require discrete relevance labels. Moreover, enforcing a single correct station per event is unrealistic in EV--EV energy trading, where multiple candidate stations may be simultaneously viable. We therefore generate \emph{event-wise graded relevance} labels using a rank-based mapping that is robust to the absolute scale of $P_{e,j}$.

For each event $e$, candidates are ranked by descending soft relevance $P_{e,j}$. Let $\mathrm{rank}_e(j)\in\{1,\ldots,|\mathcal{J}_e|\}$ denote the rank of station $j$ within event $e$, with $\mathrm{rank}_e(j)=1$ indicating the highest relevance. We define a normalized rank score
\begin{equation}
q_{e,j}
=
1 - \frac{\mathrm{rank}_e(j)-1}{\max(|\mathcal{J}_e|-1,1)},
\qquad
q_{e,j}\in[0,1].
\end{equation}
This transformation depends only on relative ordering within the event and is therefore stable even when $P_{e,j}$ values are close due to probabilistic smoothing.

We convert $q_{e,j}$ into graded relevance labels $y_{e,j}\in\{0,1,\ldots,G\}$ using ordered thresholds $\kappa_G>\kappa_{G-1}>\cdots>\kappa_1$:
\begin{equation}
y_{e,j}=
\begin{cases}
G, & q_{e,j}\ge \kappa_G,\\
G-1, & \kappa_{G-1}\le q_{e,j}<\kappa_G,\\
\vdots & \\
1, & \kappa_1\le q_{e,j}<\kappa_2,\\
0, & q_{e,j}<\kappa_1.
\end{cases}
\end{equation}
Thresholds $\{\kappa_g\}$ are selected to reflect the desired fraction of top-ranked candidates per event (e.g., top 10\%, 30\%, 60\%), ensuring label diversity and preserving the within-event ordering signal required for learning-to-rank.

 Together, these steps transform heuristic, role-aware suitability scores into supervised graded relevance labels suitable for learning-to-rank without requiring observed station selection data.

\subsection{Data preprocessing}

EV mobility and energy data undergo a preprocessing pipeline involving data understanding, data cleaning, data encoding and transformation, as well as feature extraction and selection to support charging nodes recommendation in EV--EV energy trading.

\subsubsection{Data understanding:} 
\begin{figure*}
\centering
\subfloat[Density distribution of vehicles and charging stations in Chicago at specific time intervals. The x-axis and y-axis represent the longitude and latitude of vehicles and charging stations, respectively.\label{fig:ev-soc-a}]{
\includegraphics[height=5.5cm,width=0.49\textwidth]{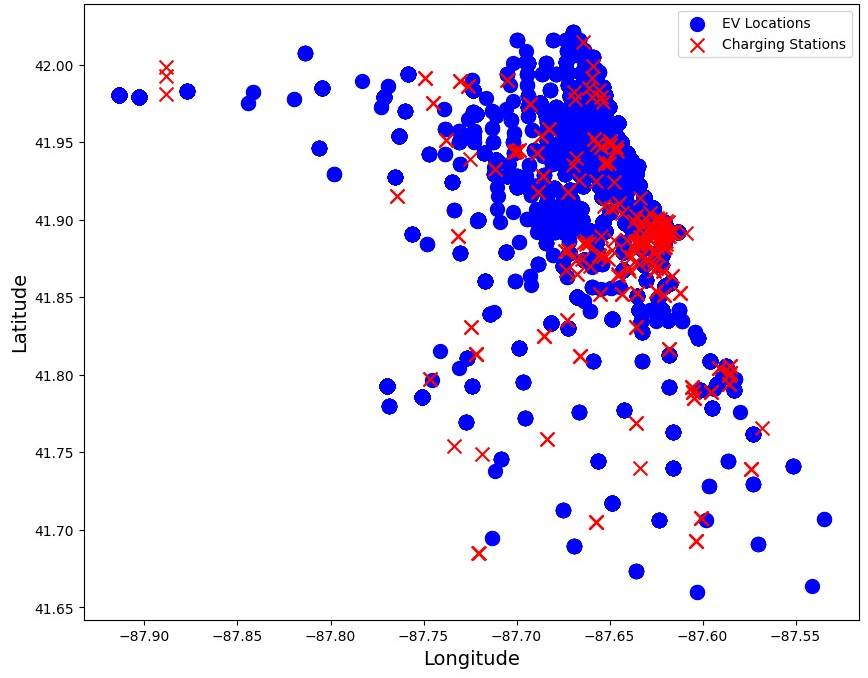}}\hfill
\subfloat[Spatial distribution of trace locations for a single EV over the year to illustrate movement patterns across different locations over a one-year period. The left-bottom inset shows the spatial distribution of trace locations for the same EV over 24 hours, illustrating movement patterns throughout a single day.\label{fig:ev-soc-b}]{
\includegraphics[height=5.5cm,width=0.49\textwidth]{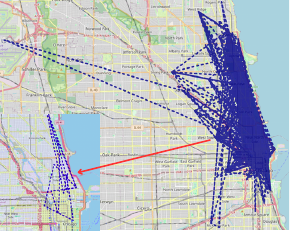}}\\
\caption{Spatial distribution of EV charging infrastructure and mobility patterns in Chicago.}
\label{fig:ev-soc-details}
\end{figure*}

The prepared dataset comprises data on taxi mobility, charging stations, and the temporal popularity of charging stations. To understand vehicle mobility, we visualized the spatial distribution of EVs and charging stations across selected time intervals (Figure \ref{fig:ev-soc-a}). This analysis provides insights into charging provider and consumer hotspots, as well as potential congestion areas across spatial and temporal dimensions. These findings are crucial for optimizing charging station distribution and enhancing energy distribution efficiency.

In addition, the spatial distribution of journey points for a single EV over a 24-hour period and across the entire year (Figure \ref{fig:ev-soc-b}) reveals recurring mobility patterns, including frequently traversed routes, repeated stop-off locations, and spatial concentration around specific urban regions. These patterns indicate that vehicle mobility is not random but exhibits regular spatiotemporal structure, with certain areas repeatedly acting as interaction hotspots where vehicles are more likely to engage in energy trading. Such recurring locations and routes are particularly relevant for EV--EV energy trading, as they represent candidate regions where supply–demand matching and charging node coordination are more likely to occur. This analysis motivates the use of historical journey patterns and spatial recurrence as informative features for recommending next charging nodes.


\subsubsection{Data cleaning}

Data cleaning was performed to address missing values, duplicate records, and data inconsistencies to improve data quality and maintain data integrity. Duplicate entries were removed when the same EV was recorded traveling between identical locations at the same timestamp. In addition, loop journeys--cases where an EV started and ended at the same location without meaningful displacement (i.e., traveled distance close to zero)--were excluded, as they provide limited relevance to EV–EV trading and add unnecessary noise. To resolve inconsistencies, measurement units (Wh, km) were standardized across all key attributes, including battery energy level, travel distance, trading volume, charging speed, and battery capacity, ensuring consistency and comparability throughout the analysis.


\subsubsection{Data encoding and transformation}

We applied appropriate encoding techniques to the dataset to ensure compatibility with ranking models. Categorical features such as EV model, candidate charging stations, and trading role were encoded using the label encoding method \citep{poslavskaya2023encoding}. 
We examined the numerical attributes in the dataset and observed substantial scale variations across several features, including \emph{distance\_to\_cs}, \emph{charging\_speed}, \emph{volume\_of\_energy}, and \emph{popularity\_index}. Such disparities in magnitude can bias machine learning models for learning-to-rank by causing features with larger numeric ranges to dominate the learning process. To address this issue, min–max normalization method \citep{hastie2009elements} was applied to these continuous attributes, ensuring that all features were scaled to a comparable range while preserving their original distributions.

\subsubsection{Feature extraction and selection}

To support next charging node recommendation under uncertainty, we construct a multidimensional feature set that captures candidate station characteristics, EV state information, and temporal and contextual factors available at each decision point. Feature extraction is performed at the level of EV decision events, where each event corresponds to a set of candidate charging stations considered simultaneously. Table~\ref{tab:features} summarizes the features used for model training and evaluation. Identifiers such as EV ID and EV journey ID are retained only for query grouping, data splitting, and evaluation purposes and are not used as input features during model training. Similarly, candidate charging node identifiers are not used as model input features. The learning-to-rank model operates on candidate-level attributes and EV contextual features to produce relevance scores for each candidate within a decision event. Station identifiers are used only to associate predicted relevance scores with physical charging nodes after ranking.

\begin{table*}
\scriptsize
\centering
\caption{Feature set used for learning-to-rank in EV--EV charging node recommendation.The following EV-level, temporal, spatial, context-level, and vehicle-level attributes are treated as contextual features, as they characterize the decision environment under which candidate charging nodes are ranked.}
\label{tab:features}
\renewcommand{\arraystretch}{1.15}
\begin{tabular}{p{3.6cm} p{4.2cm} p{6.2cm}}
\hline
\textbf{Category} & \textbf{Feature} & \textbf{Description} \\
\hline

\multirow{3}{*}{Candidate-level} 
& Distance to station 
& Distance from the EV’s current location to the candidate charging station, capturing travel effort and feasibility. \\ \cline{2-3}

& Charging speed 
& Maximum charging or discharging rate supported by the station, reflecting energy transfer capability. \\ \cline{2-3}

& Station popularity 
& Historical usage intensity of the station, reflecting recurring demand and supply pressure and local availability conditions that influence EV--EV matching feasibility. \\

\hline
\multirow{6}{*}{EV-level}
& Battery capacity 
& Maximum battery capacity of the EV. \\ \cline{2-3}

& Battery energy level 
& Available battery energy at the decision point. \\ \cline{2-3}

& State-of-charge ($\mathrm{SoC}_e$) 
& Normalized battery energy level, capturing charging or discharging urgency. \\ \cline{2-3}

& Transaction pressure ($p_e$) 
& Role-aware pressure reflecting supply or demand urgency in EV--EV energy trading. \\ \cline{2-3}

& Energy quantity 
& Amount of energy requested or offered for trading, depending on the EV’s role. \\ \cline{2-3}

& Trading role 
& EV role in the trading interaction (provider or consumer). \\

\hline
\multirow{3}{*}{Temporal}
& Time-of-day 
& Hour and minute of the decision point, capturing daily mobility and charging patterns. \\ \cline{2-3}

& Calendar context 
& Day-of-week, week, and month indicators capturing periodic mobility behavior. \\ \cline{2-3}

& Cyclic time encodings 
& Sine and cosine transformations of time-of-day, day-of-week, and month to model temporal periodicity. \\

\hline
\multirow{2}{*}{Spatial}
& Location identifier 
& Encoded representation of the EV’s current, source, and destination locations at the decision point. \\ \cline{2-3}

& Community area 
& Encoded spatial region providing coarse-grained geographic context. \\

\hline
Context-level
& Candidate set size 
& Number of candidate charging stations available for the EV decision event, capturing choice complexity and infrastructure density. \\

\hline
Vehicle-level
& EV model 
& Vehicle model type capturing heterogeneity in charging and discharging characteristics. \\

\hline
\end{tabular}
\end{table*}

\begin{figure}
\centering
\small 
\includegraphics[width=0.8\textwidth]{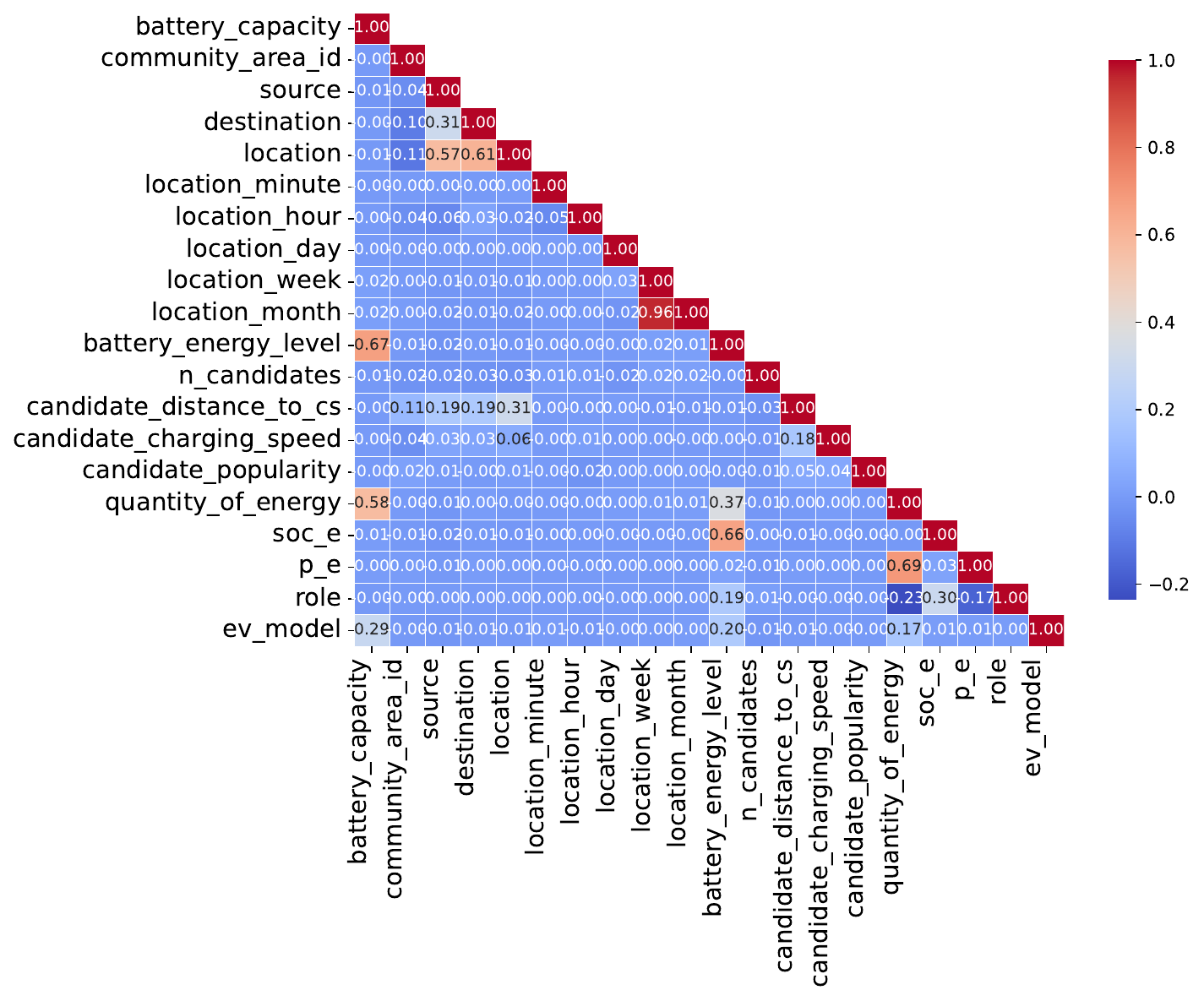}
\caption{Pearson Correlation Analysis among features: This analysis measures the linear relationships among features, identifying potential multicollinearity and redundant information.}
\label{fig:pca-corr1}
\end{figure}

To assess potential redundancy among features, we conduct a pairwise Pearson correlation analysis \citep{jebli2021prediction, mehedi2026dysec}.
Figure~\ref{fig:pca-corr1} presents the correlation matrix for the selected candidate-level, EV-level, and contextual features. Overall, correlation values remain low to moderate, with no feature pairs exhibiting strong linear dependence indicative of multicollinearity. Expected relationships are observed, such as a moderate positive correlation between battery energy level and normalized state-of-charge (\textit{soc\_e}), reflecting their shared physical meaning. The requested transaction quantity shows a positive correlation with both state-of-charge and instantaneous battery energy level, indicating that EVs with greater available energy tend to engage in larger energy trading volumes. Candidate station attributes, including distance to station, charging speed, and historical popularity, exhibit low mutual correlation, suggesting that they capture complementary aspects of station suitability rather than redundant information. Based on this analysis, no features were removed due to redundancy, and the complete feature set was retained for subsequent learning-to-rank experiments.

\subsection{Learning-to-Rank models for next charging nodes recommendation}

To generalize beyond rule-based suitability estimation and learn a unified decision policy applicable to unseen events, we formulate next charging node recommendation as a supervised learning-to-rank problem. The objective is to learn a scoring function that orders candidate stations according to their relative relevance within each decision event. Each EV decision event $e$ is treated as a query, and the corresponding candidate stations $j \in \mathcal{J}_e$ are treated as items to be ranked. For each query--item pair, the input feature vector includes candidate-specific attributes (e.g., distance, charging speed, and popularity), EV-level attributes (e.g., trading role and energy level), and optional contextual features (e.g., candidate count) that are shared across all candidates within the same event.
The target supervision signal is the graded relevance label $y_{e,j}$. These labels indicate relative station desirability within an event and do not represent observed station selection outcomes.

To learn this ranking function, we employ supervised learning-to-rank models trained using graded relevance labels derived from probabilistically refined station suitability scores. This formulation captures relative station desirability under uncertainty without assuming a single ground-truth choice. Specifically, we evaluate three gradient-boosting-based ranking algorithms: LightGBM Ranker with a LambdaRank objective, XGBoost Ranker with a pairwise ranking objective, and CatBoost Ranker with a PairLogit objective \citep{sun2024multi, jeunen2023gradient}. All models are trained independently on the NextTrade-EV dataset using identical feature sets and relevance labels, enabling a fair comparison of ranking performance.

Given a set of queries and associated graded relevance labels, the ranking models learn a scoring function $f(\cdot)$ that assigns higher scores to stations with higher relevance grades within the same decision event. Depending on the underlying algorithm, pairwise or listwise loss functions are employed to penalize incorrect orderings, with larger penalties assigned to swaps involving greater relevance differences. As the relevance labels are inferred rather than directly observed, the resulting models should be interpreted as learning a generalized ranking policy consistent with inferred station suitability patterns, rather than predicting realized driver choices.

\begin{figure}
\centering
\small 
\includegraphics[width=0.8\textwidth]{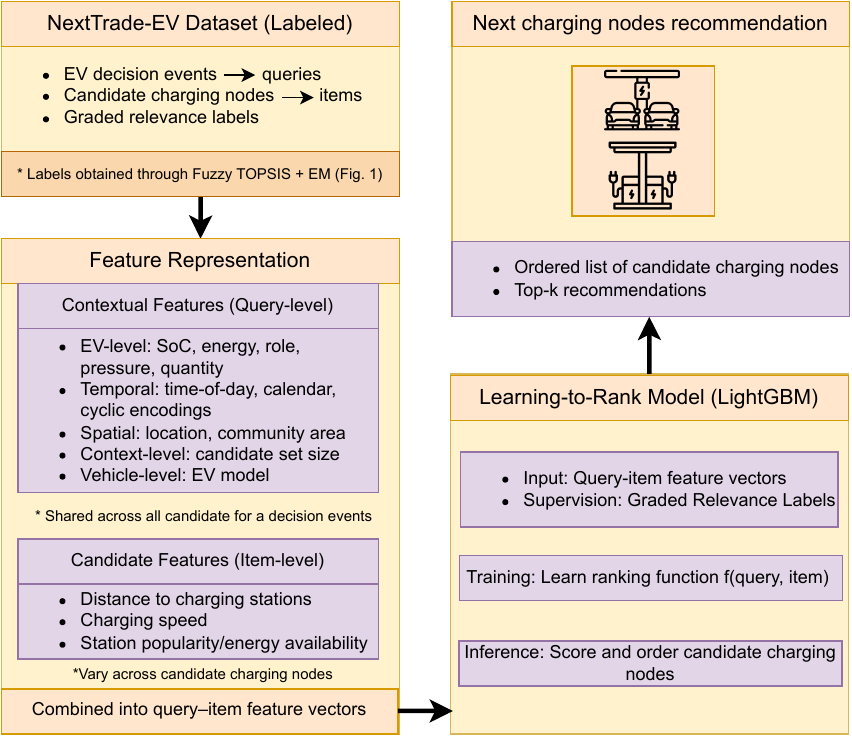}
\centering
\caption{Next charging nodes recommendation framework based on learning-to-rank.}
\label{fig:framework-1}
\end{figure}

In summary, the proposed approach formulates each EV decision event as a ranking query, combines contextual and candidate-level attributes into query--item feature vectors, and learns a ranking function using a gradient-boosted learning-to-rank model. The complete learning-to-rank–based recommendation framework is illustrated in Fig. \ref{fig:framework-1}.

\subsection{Model training}

\begin{figure}
\centering
\includegraphics[width=0.8\textwidth]{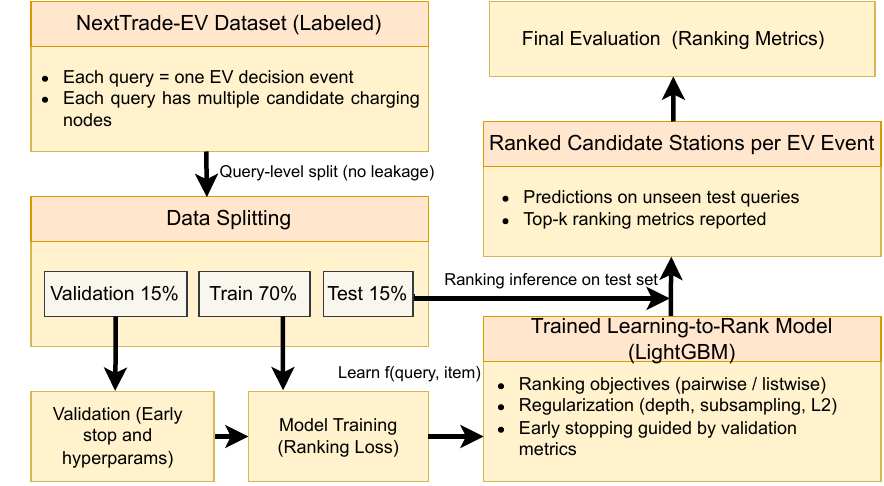}
\caption{Query-level training, validation, and testing protocol for learning-to-rank models.}
\label{fig:protocol}
\end{figure}

The dataset was partitioned into training (70\%), validation (15\%), and test (15\%) sets at the \emph{query level}, where each query corresponds to an EV decision event. This splitting strategy ensures that all candidate stations associated with the same event are assigned to a single split, thereby preventing information leakage across training, validation, and test sets. 
The validation set was used exclusively for hyperparameter tuning and early stopping, while the test set was reserved for final performance evaluation. All ranking models were trained using their respective ranking objectives and optimized to preserve the relative ordering of candidate stations within each event, as illustrated in Fig. \ref{fig:protocol}.


To mitigate overfitting, model-specific regularization techniques inherent to gradient-boosted decision trees were employed, including tree depth constraints, subsampling of instances and features, and $\ell_2$ regularization on leaf weights. Early stopping based on validation ranking metrics was applied where supported. These measures ensure robust generalization to unseen EV decision events.

To further assess robustness, we additionally conducted query-level $K$-fold cross-validation, where EV decision events were randomly partitioned into $K$ disjoint folds and all candidate stations associated with the same event were kept within a single fold. Models were trained on $K-1$ folds and evaluated on the remaining fold, and performance was averaged across folds. The resulting trends were consistent with those obtained using the fixed train/validation/test split.

\subsection{Performance evaluation metrics}

To evaluate the quality of next charging node recommendation, we employ standard ranking metrics that assess the relative ordering of candidate stations within each EV decision event. Since the task is formulated as a ranking problem with graded relevance labels rather than a single ground-truth choice, we focus on top-$k$ ranking metrics that emphasize early precision and coverage.
We report ranking performance at cutoffs $k \in \{1,3,5,10\}$, which reflect realistic recommendation list sizes in EV--EV energy trading scenarios.

\textbf{NDCG@$k$:}
Normalized Discounted Cumulative Gain (NDCG) evaluates ranking quality under graded relevance by rewarding highly relevant stations appearing near the top of the ranked list. The Discounted Cumulative Gain at cutoff $k$ is defined as
\begin{equation}
\mathrm{DCG}@k = \sum_{i=1}^{k} \frac{2^{\mathrm{rel}_i}-1}{\log_2(i+1)},
\end{equation}
where $\mathrm{rel}_i$ denotes the graded relevance of the station ranked at position $i$. NDCG is obtained by normalizing DCG with respect to the ideal ranking:
\begin{equation}
\mathrm{NDCG}@k = \frac{\mathrm{DCG}@k}{\mathrm{IDCG}@k},
\end{equation}
where $\mathrm{IDCG}@k$ is the DCG of the ideal (perfectly ordered) ranking. NDCG is particularly suitable for our setting as it supports multi-level relevance labels derived from probabilistic smoothing \citep{burges2010ranknet}.

\textbf{Recall@$k$:}
Recall@$k$ measures the extent to which relevant stations are covered within the top-$k$ ranked results. Let $\mathcal{R}_e$ denote the set of relevant stations for event $e$, defined as those with graded relevance above a predefined threshold, and let $\hat{\mathcal{R}}_{e,k}$ denote the top-$k$ ranked stations. Recall@$k$ is computed as
\begin{equation}
\mathrm{Recall}@k = \frac{|\mathcal{R}_e \cap \hat{\mathcal{R}}_{e,k}|}{|\mathcal{R}_e|}.
\end{equation}
This metric reflects the ability of the model to include plausible charging or trading options within a limited recommendation list. In scenarios where only one station is highly relevant, Recall@$k$ reduces to a hit-rate indicator \citep{sun2024multi}.

\textbf{MRR:}
Mean Reciprocal Rank (MRR) emphasizes how early the first relevant station appears in the ranked list. It is defined as
\begin{equation}
\mathrm{MRR} = \frac{1}{Q} \sum_{e=1}^{Q} \frac{1}{\mathrm{rank}_e},
\end{equation}
where $\mathrm{rank}_e$ denotes the rank position of the first station in event $e$ whose relevance exceeds the relevance threshold, and $Q$ is the total number of events. Higher MRR values indicate that highly suitable stations tend to appear closer to the top of the recommendation list \citep{burges2010ranknet}.

\section{Result analysis and discussions}
\label{rad}
This section evaluates the overall performance of the selected models through a comprehensive analysis. 

\subsection{Experimental specifications}

For scalable training and evaluation, we executed experiments on a high-performance computing (HPC) cluster using the Portable Batch System (PBS). Each run used a single compute node provisioned with a 4-core CPU, 512\,GB RAM, and one NVIDIA H100 GPU. The software stack included Python~3.10.8 with Pandas, NumPy, Scikit-learn, TensorFlow, and Matplotlib to support data processing and model training. Jobs were submitted via a PBS script and consolidated stdout/stderr logging to \emph{data.log}. Resource requests and environment activation were handled at submission time to ensure reproducibility and efficient utilization of GPU resources.

\subsection{Performance analysis}

Based on the NextTrade-EV datasets, each decision point produces a probabilistic distribution of candidate charging stations that represents the relative likelihood of selection.
These soft labels are used to train ranking models that aim to learn an ordering function over candidate stations at each decision point in a journey for the EV driver.
To evaluate their ranking effectiveness, we adopt standard information-retrieval metrics--NDCG@k, Recall@k, and MRR--which measure how highly relevant station is ranked among the top-$k$ candidates.
Unlike accuracy-based evaluation, these metrics emphasize the quality of the entire ranked list, offering a more fine-grained assessment of model behavior at each decision point.

\begin{table*}
\scriptsize
\centering
\caption{Comparison of learning-to-rank models for next charging nodes recommendation}
\label{tab:ltr_comparison}
\begin{tabular}{lccccccccc}
\hline
\textbf{Model} 
& \textbf{NDCG@1} 
& \textbf{Recall@1} 
& \textbf{NDCG@3} 
& \textbf{Recall@3} 
& \textbf{NDCG@5} 
& \textbf{Recall@5} 
& \textbf{NDCG@10} 
& \textbf{Recall@10} 
& \textbf{MRR} \\
\hline
LightGBM (LambdaRank)
& \textbf{0.9795} & \textbf{0.1931}
& \textbf{0.9725} & 0.5112
& \textbf{0.9762} & 0.6848
& \textbf{0.9811} & 0.8889
& \textbf{0.9990} \\

XGBoost (Pairwise)
& 0.9655 & 0.1926
& 0.9623 & \textbf{0.5119}
& 0.9689 & \textbf{0.6854}
& 0.9749 & 0.8899
& 0.9984 \\

CatBoost (PairLogit)
& 0.9678 & 0.1924
& 0.9646 & 0.5111
& 0.9709 & \textbf{0.6854}
& 0.9772 & \textbf{0.8900}
& 0.9982 \\
\hline
\end{tabular}
\end{table*}

Table~\ref{tab:ltr_comparison} represents the ranking performance of three supervised learning-to-rank model including LightGBM (LambdaRank), XGBoost (pairwise), and CatBoost (PairLogit), evaluated on the NextTrade-EV dataset (test data) using graded relevance labels derived from probabilistically smoothed station suitability scores. All models were trained and evaluated under identical query-level data splits and feature sets to ensure a fair comparison.

Across all evaluation metrics and cutoff levels ($k\in{1,3,5,10}$), LightGBM consistently achieves the best overall performance. In particular, LightGBM attains the highest NDCG@1 (0.9795), indicating superior accuracy in placing the most relevant charging or trading station at the top of the recommendation list. This advantage persists at larger cutoffs, with LightGBM maintaining the highest NDCG@3, NDCG@5, and NDCG@10 values, demonstrating strong ranking quality throughout the recommendation list rather than only at the top position.

\begin{figure}
\centering
\small 
\includegraphics[width=0.8\textwidth]{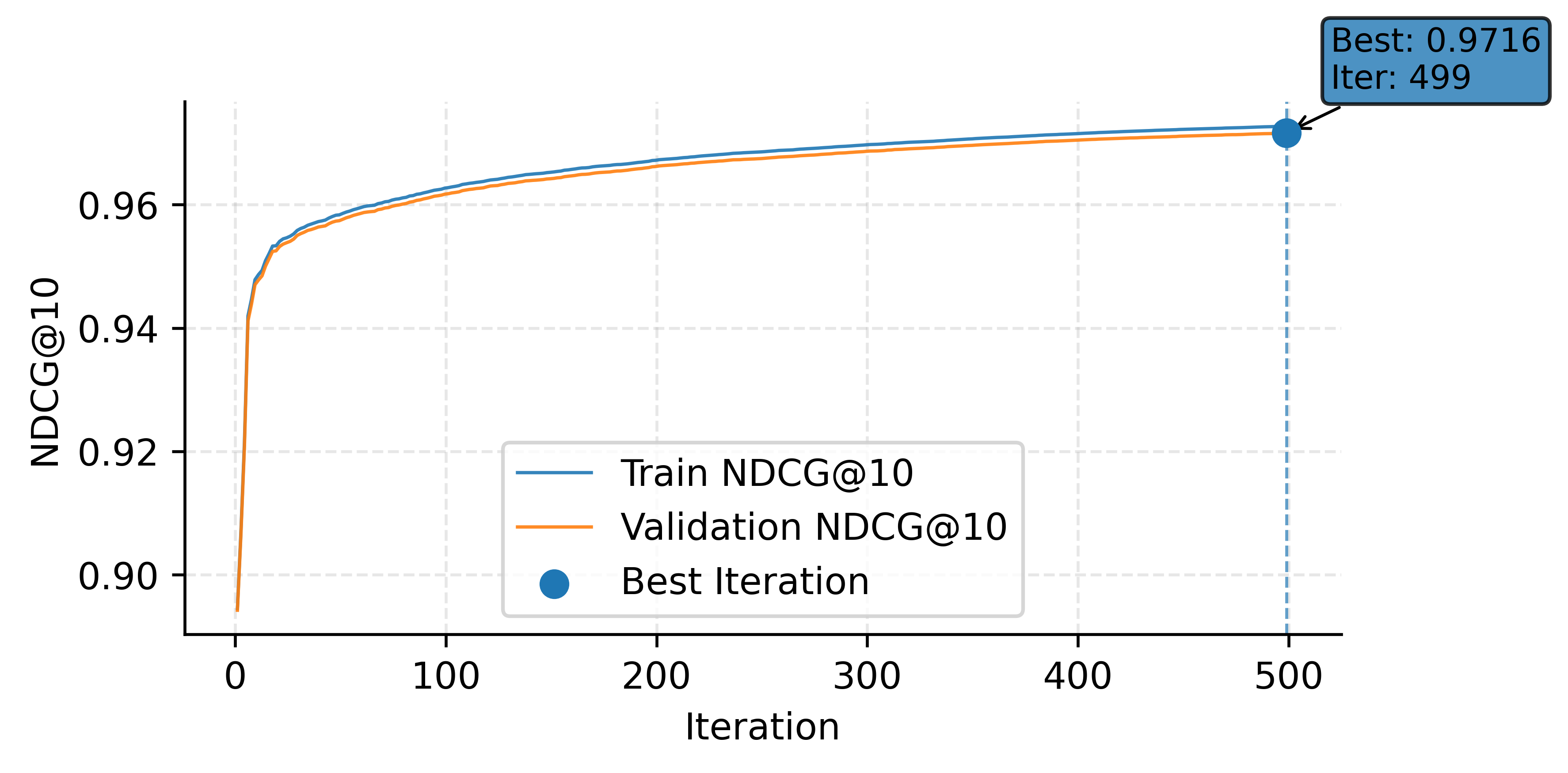}
\centering
\caption{LightGBM LambdaRank convergence and early stopping. Training (blue) and validation (purple) NDCG@10 versus boosting iteration for the LightGBM ranker. The best iteration selected by early stopping occurs at Iter = 499 with a peak validation NDCG@10 = 0.9716 (blue marker; dashed vertical line).}
\label{fig:lgb_convergence}
\end{figure}

Recall-based metrics reveal a closely competitive pattern. Recall@1 is nearly identical across models, with LightGBM achieving a marginal advantage (0.1931). Recall trends are broadly consistent across models: XGBoost is marginally higher at Recall@3 and Recall@5, whereas CatBoost attains the best Recall@10, with all differences remaining small. This indicates that while XGBoost and CatBoost retrieve slightly more relevant stations at specific deeper cutoffs, LightGBM achieves superior ranking quality, as reflected by consistently higher NDCG scores. Hence, the results suggest that LightGBM prioritizes more accurate ordering of highly relevant stations at top positions, while differences in recall at deeper cutoffs remain marginal across models.

Mean Reciprocal Rank (MRR) values are uniformly high for all models, exceeding 0.997, which indicates that at least one highly relevant station typically appears near the top of the ranked list for most decision events. Nonetheless, LightGBM again achieves the highest MRR (0.9990), reinforcing its superior early-ranking behavior. 

Figure~\ref{fig:lgb_convergence} illustrates the convergence behavior of LightGBM LambdaRank. The training and validation NDCG@10 curves follow a consistent trajectory and stabilize after sufficient boosting iterations (around 500 iterations), with minimal divergence between them. Early stopping identifies iteration 499 as the optimal model based on validation NDCG@10. After convergence, additional boosting iterations produce only marginal changes in performance, confirming stable optimization behavior.

These results indicate that gradient-boosted decision tree–based ranking models are well suited to the EV–EV charging node recommendation problem. Among the evaluated models, LightGBM provides the best balance between early precision and overall ranking consistency and is therefore adopted as the primary ranker in the proposed framework. For subsequent analysis, LightGBM is fixed as the ranking model, while TOPSIS is used as the baseline supervision mechanism, allowing the effect of probabilistic refinement to be isolated independently of model architecture.

\subsection{Sensitivity analysis}
To examine the robustness of the proposed EV–EV trading recommendation framework, we conduct sensitivity analyses on key design parameters that influence role assignment and candidate generation. Specifically, we evaluate the impact of (1) SoC threshold selection used in journey-level role assignment and (2) spatial radius constraints used for candidate charging pad generation. These analyses assess whether ranking performance remains stable under reasonable variations of operational and spatial assumptions.
\subsubsection{Sensitivity to SoC threshold}
\label{sensitivity-radius-soc}

To assess the robustness of the proposed mechanism, we conducted a sensitivity analysis on the surplus thresholds used in Algorithm~\ref{alg:energy_volume_role}. The lower cutoff ($r \le 0.30$) corresponds to a minimum operational safety constraint, ensuring that EVs with insufficient surplus are deterministically classified as consumers. Accordingly, this boundary was kept fixed. Hence, sensitivity analysis focuses on the upper provider cutoff, which governs the strictness of surplus interpretation. Specifically, we varied the provider threshold around the baseline configuration $(r \le 0.30, r > 0.90)$ by testing alternative values of $r > 0.85$ and $r > 0.95$. The probabilistic decision rule, including the probability parameters $p_{\text{mid}}$ and $p_{\text{high}}$, was preserved, and a fixed random seed was employed during probabilistic role assignment to ensure reproducibility. This setup evaluates whether ranking performance is sensitive to reasonable perturbations of the deterministic provider threshold.

\begin{table*}

\centering
\caption{Sensitivity Analysis of SoC Thresholds}
\label{tab:soc_sensitivity}
\scriptsize
\begin{tabular}{lccccccccc}
\hline
\textbf{Threshold Configuration ($r$)} & \textbf{Providers (\%)} & \textbf{Consumers (\%)} & \textbf{NDCG@1} & \textbf{Recall@1}& \textbf{NDCG@5} & \textbf{Recall@5} & \textbf{NDCG@10} & \textbf{Recall@10} \\
\hline
$(r \le 0.30,\; r > 0.85)$ & 55.99 & 44.01 & 0.9793 & 0.1930 & 0.9761 & 0.6849 & 0.9811 & 0.8889 \\ \hline
\textbf{Baseline $(r \le 0.30,\; r > 0.90)$} & 56.48 & 43.52 &  0.9795 & 0.1931 & 0.9762 &0.6848 & 0.9811 & 0.8889 \\ \hline
$(r \le 0.30,\; r > 0.95)$ & 56.78 & 43.22 & 0.9795 & 0.1931 & 0.9763 & 0.6849 & 0.9812  & 0.8889 \\
\hline
\end{tabular}
\end{table*}

Table \ref{tab:soc_sensitivity} presents the sensitivity analysis of the provider cutoff while keeping the consumer safety threshold fixed at $r \le 0.30$. Varying the upper surplus threshold from 0.85 to 0.95 results in only minor changes in role (provider, consumer) distribution, with the proportion of providers ranging between 55.99\% and 56.78\%. This indicates that moderate adjustments to the provider cutoff do not substantially alter the class balance.

More importantly, ranking performance remains unchanged across configurations. NDCG@1 remains at approximately 0.9795, NDCG@5 at 0.9762, and NDCG@10 at 0.9811, while Recall@10 consistently stays around 0.8889. The negligible variation across all ranking depths demonstrates that the learned relevance function is not sensitive to moderate perturbations of the SoC threshold.

These findings confirm that the proposed role-assignment mechanism and ranking model are robust with respect to reasonable variations in the provider surplus cutoff, and that performance does not depend critically on the exact selection of the 90\% threshold.

\subsubsection{Sensitivity to candidate radius}
\label{sensitivity-radius}

\begin{figure}
\centering
\small 
\includegraphics[width=0.8\textwidth]{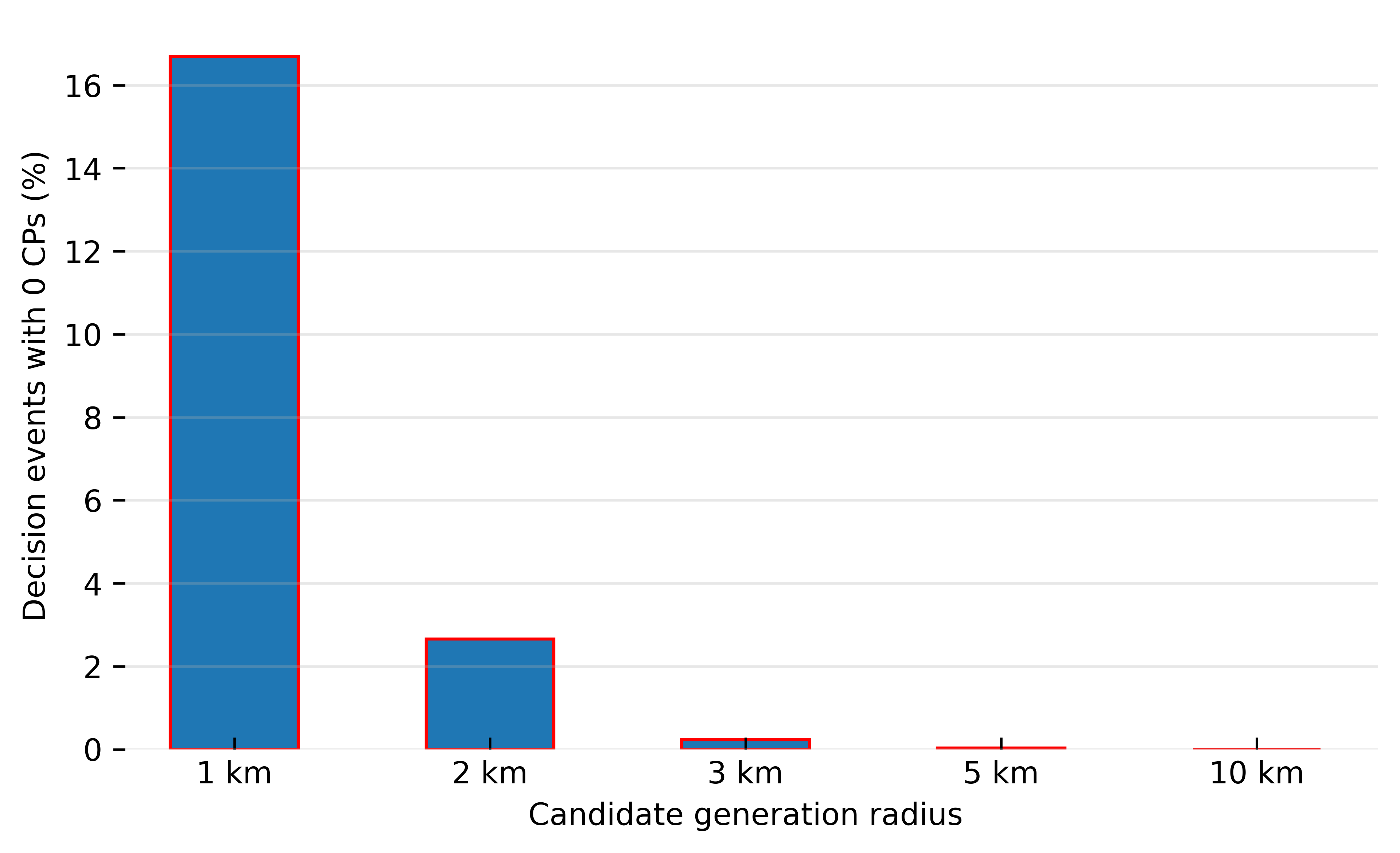}
\centering
\caption{Percentage of decision events with zero candidate WCPs under varying geographic filtering radius (1–10 km). The proportion of infeasible events decreases sharply as the radius expands, dropping from 16.84\% at 1 km to nearly zero beyond 5 km. This demonstrates that small radius can lead to substantial candidate sparsity, whereas radius $\geq$3 km effectively ensure candidate availability for almost all decision events.}
\label{fig:radius_statistics}
\end{figure}

\begin{figure}
\centering
\small 
\includegraphics[width=0.8\textwidth]{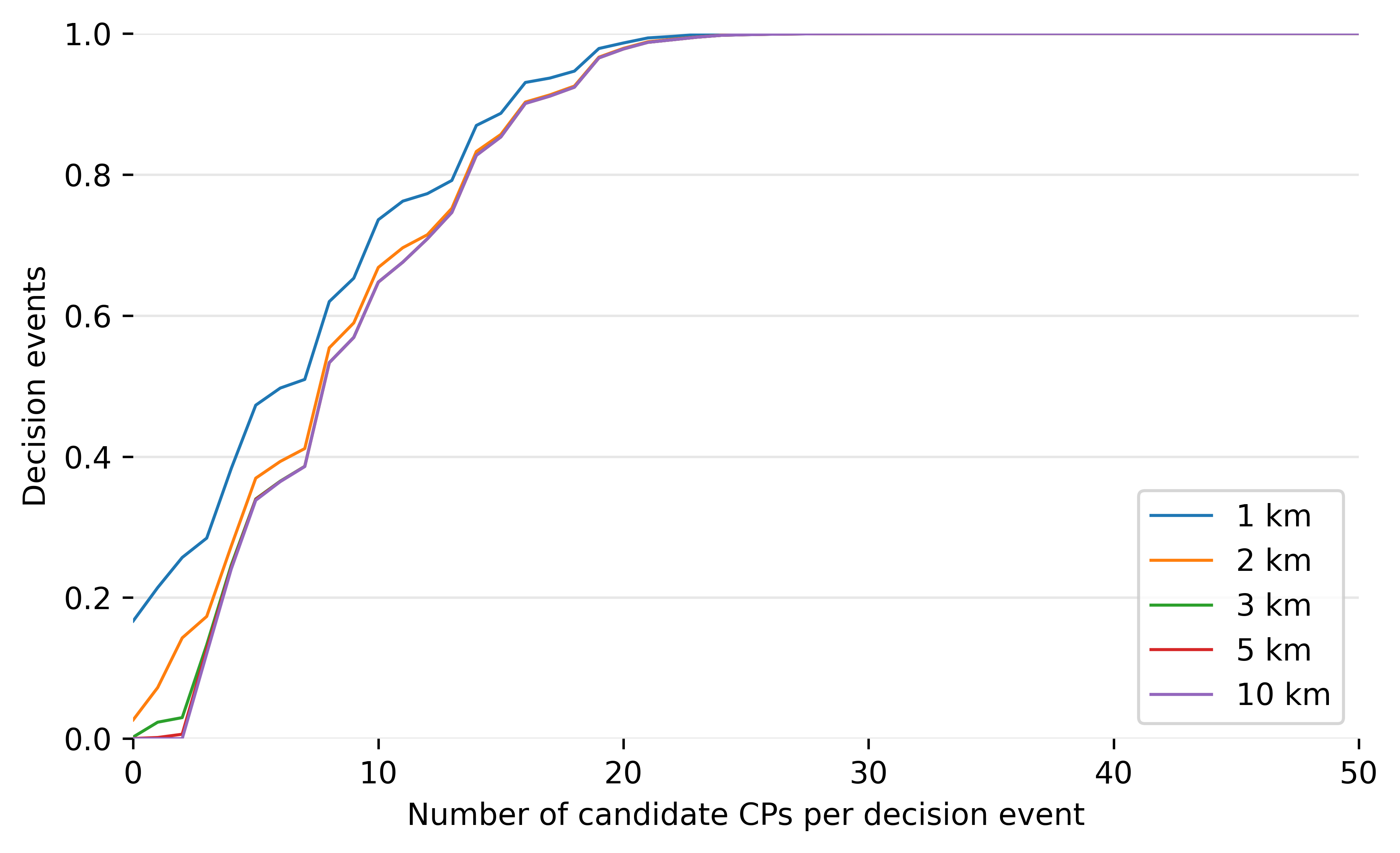}
\centering
\caption{The empirical cumulative distribution of candidate charging pads per decision event under varying radius limits. For a given value $x$, the cumulative distribution represents the proportion of decision events with at most $x$ candidate WCPs. The curves demonstrate that candidate availability stabilizes beyond 3\,km, as the distributions for 5\,km and 
10\,km are nearly indistinguishable.}
\label{fig:radius-cdf}
\end{figure}

To evaluate the robustness of our model with respect to the candidate generation radius, we varied the maximum allowable detour distance from 1\,km to 10\,km. Table \ref{tab:cp_distribution_radius} summarizes the distribution of candidate WCPs per decision event across radius of 1, 2, 3, 5, and 10 km over 2190331 events. The mean number of candidates increases from 7.21 at 1 km to 9.15 at 10 km, while the median stabilizes at 8 for radius $\geq$ 2 km, indicating that most accessible WCPs are already captured within a short spatial range. The upper percentiles remain constant (P90 = 16, P95 = 19) across radius, showing that expanding the radius does not substantially increase the upper tail of candidate availability. The primary effect of increasing the radius is the reduction of infeasible events: the proportion of zero-candidate events drops sharply from 16.84\% at 1 km to 0.23\% at 3 km and becomes negligible beyond 5 km. Similarly, sparse cases ($\leq$ 3 candidates) decrease from 28.72\% at 1 km to 13.85\% at 3 km, while the proportion of events with $\leq$ 10 candidates decreases more moderately, reflecting a rightward shift of the distribution rather than explosive candidate growth. These findings indicate that increasing the radius beyond 3 km yields diminishing returns in candidate expansion. A 3 km radius already eliminates nearly all infeasible events while maintaining moderate candidate set sizes, suggesting that most practically accessible charging options are captured within this spatial range. These observations are also visually supported by Figures \ref{fig:radius_statistics} and \ref{fig:radius-cdf}.

\begin{table*}
\centering
\footnotesize
\caption{Distribution of candidate WCPs per EV decision event under varying geographic radius limits.}
\label{tab:cp_distribution_radius}
\begin{tabular}{c c c c c c c c c c}
\hline
\textbf{Radius (km)} & \textbf{\#Events} & \textbf{Mean} & \textbf{Median} & \textbf{P90} & \textbf{P95} & \textbf{0 WCPs (\%)} & $\boldsymbol{\leq 3}$ (\%) & $\boldsymbol{\leq 5}$ (\%) & $\boldsymbol{\leq 10}$ (\%) \\
\hline
1   & 2190331 & 7.21 & 6 & 16 & 18 & 16.84 & 28.72 & 48.31 & 74.80 \\
2   & 2190331 & 8.64 & 8 & 16 & 19 & 2.67  & 17.75 & 38.16 & 68.32 \\
3   & 2190331 & 9.07 & 8 & 16 & 19 & 0.23  & 13.85 & 35.33 & 66.34 \\
5   & 2190331 & 9.14 & 8 & 16 & 19 & 0.02  & 13.18 & 35.27 & 66.33 \\
10  & 2190331 & 9.15 & 8 & 16 & 19 & 0.00  & 12.54 & 35.15 & 66.32 \\
\hline
\end{tabular}
\end{table*}

When examined alongside the ranking results in Table~\ref{tab:radius_sensitivity}, a consistent pattern emerges. Despite differences in candidate sparsity between the 1 km radius and larger radii, ranking performance remains highly stable. At 1 km, NDCG@1 is 0.9764 and Recall@1 is 0.2212, while at 10 km they are 0.9796 and 0.1930, respectively. Similarly, NDCG@10 varies only slightly from 0.9782 (1 km) to 0.9813 (10 km), and Recall@10 remains close to 0.89 across all radius settings. Although 74.80\% of 1 km decision events contain ten or fewer candidates, compared to approximately 66.3\% at 3 km and above, the ranking model maintains comparable accuracy. This indicates that the model does not rely on trivially small candidate sets to achieve strong performance. Under the 1 km setting, candidate availability is more restricted and some events exhibit sparsity, whereas under the 10 km setting the model must discriminate among a larger and more diverse set of charging options. The sustained ranking quality under expanded search spaces demonstrates that the learned relevance function effectively captures mobility context, charging pad features, and energy context, rather than benefiting from artificially constrained candidate pools. These findings demonstrate the robustness of the proposed framework with respect to spatial filtering parameters.

\begin{table*}
\centering
\footnotesize
\caption{Sensitivity of ranking performance to candidate-generation radius.}
\label{tab:radius_sensitivity}
\begin{tabular}{c c c c c c c c c c}
\hline
\textbf{Radius (km)} &
\textbf{NDCG@1} & \textbf{Recall@1} &
\textbf{NDCG@3} & \textbf{Recall@3} &
\textbf{NDCG@5} & \textbf{Recall@5} &
\textbf{NDCG@10} & \textbf{Recall@10} \\
\hline
1 & 0.9764 & 0.2212 & 0.9693 & 0.5160 & 0.9726 & 0.6866 & 0.9782 & 0.8862 \\
2 & 0.9750 & 0.2442 & 0.9668 & 0.5164 & 0.9704 & 0.6840 & 0.9765 & 0.8828 \\
3 & 0.9768 & 0.2040 & 0.9679 & 0.5115 & 0.9716 & 0.6817 & 0.9782 & 0.8848 \\
5 & 0.9796 & 0.1959 & 0.9705 & 0.5143 & 0.9746 & 0.6854 & 0.9813 & 0.8886 \\
10 & 0.9796 & 0.1930 & 0.9706 & 0.5123 & 0.9747 & 0.6842 & 0.9813 & 0.8881 \\
\hline
\end{tabular}
\end{table*}

In summary, the sensitivity analyses across both surplus threshold configurations and spatial candidate-generation radius demonstrate that the proposed framework exhibits stable ranking behavior under reasonable perturbations of key design parameters. The negligible variation in NDCG and Recall metrics indicates that the learned relevance function captures structural relationships between mobility context, energy context, and charging station suitability, rather than relying on finely tuned thresholds or constrained candidate pools.

\subsection{Ablation study of probabilistic smoothing and feature components}
\label{sec:ablation}


Table \ref{tab:ablation_study} summarizes the ranking performance of the proposed framework under different ablation settings, isolating the effects of probabilistic smoothing and feature composition. We conduct this ablation study to examine the contribution of two factors in the proposed EV–EV charging node recommendation framework: (i) probabilistic smoothing through the Expectation–Maximization (EM) algorithm and (ii) the inclusion of EV-level and contextual features beyond candidate WCP attributes. All variants are trained and evaluated using the same LightGBM ranking model with a LambdaRank objective, identical query-level data splits, and the same evaluation metrics to ensure a fair comparison. The only differences across variants are the relevance label construction strategy, that is, TOPSIS-derived or EM-refined labels, and the feature set used for training.

\textbf{TOPSIS without EM:}
As a baseline, fuzzy-weighted TOPSIS suitability scores $r_{e,j}$ are directly converted into graded relevance labels using a within-event rank-based mapping, without probabilistic smoothing. This configuration represents a purely heuristic ranking pipeline. The results show strong performance at higher cutoffs, particularly for Recall@5 and Recall@10, reflecting the deterministic nature of TOPSIS and its tendency to produce sharply peaked rankings.
However, this approach implicitly assumes a decisive ordering among candidate stations and does not account for uncertainty arising from dense infrastructure or the presence of multiple simultaneously viable trading options, which are common in EV--EV energy trading scenarios.

\textbf{TOPSIS with candidate-only features:}
In this ablation, the learning-to-rank model is trained using only candidate-level features (distance, charging speed, and popularity), excluding EV state variables and contextual information. Performance degrades consistently across all NDCG cutoffs compared to the full model, indicating that candidate attributes alone are insufficient to explain relative station suitability. This result highlights the importance of incorporating EV-specific factors, such as energy state and transaction pressure, as well as temporal context, in order to capture realistic decision behavior.

\textbf{EM with candidate-only features:}
Next, EM-based probabilistic smoothing is applied to the TOPSIS scores to generate soft relevance labels, while still restricting the feature set to candidate-only attributes. Performance  decreases relative to the heuristic TOPSIS baseline, particularly at early cutoffs. This behavior is expected: EM intentionally reduces extreme dominance in suitability scores in order to model decision uncertainty. When downstream learning capacity is limited by a reduced feature set, this uncertainty cannot be effectively resolved, leading to weaker early-ranking accuracy. Importantly, this result demonstrates that EM is not intended to improve heuristic rankings in isolation, but rather to provide uncertainty-aware supervision for supervised learning-to-rank models.

\textbf{Full model (EM with full feature set):}
The full model combines EM-based probabilistic smoothing with a comprehensive feature set including candidate attributes, EV state variables (e.g., state of charge and transaction pressure), as well as temporal and contextual features. This configuration achieves the strongest overall performance across all NDCG and MRR metrics, with particularly large gains at early cutoffs such as NDCG@1.

In terms of recall, the results reveal a complementary pattern. While the heuristic TOPSIS baseline attains very high Recall@5 and Recall@10, the EM-based full model maintains competitive recall while emphasizing more accurate top-ranked ordering (as reflected by substantially higher NDCG@1 and MRR values). This indicates that the learning-to-rank framework prioritizes placing the most suitable charging pad at the highest ranks rather than maximizing the total number of relevant candidates retrieved within broader cutoffs.

Hence, the ablation study demonstrates that EM-based probabilistic smoothing, when combined with contextual feature modeling, significantly improves early-ranking precision and top-position accuracy. Although heuristic approaches may retrieve a larger number of relevant candidates at deeper cutoffs, the full model provides best ranking quality at the highest positions. Particularly, this is  important in the EV–EV trading context, where an EV driver typically selects a single charging pad, and the correctness of the top-ranked recommendation directly impacts operational efficiency and user satisfaction.


\begin{table*}
\scriptsize
\centering
\caption{Ablation study evaluating the impact of probabilistic smoothing (EM) and feature components on ranking performance using LightGBM (LambdaRank).}
\label{tab:ablation_study}
\begin{tabular}{p{3.5cm}|c|c|cccc|cccc|c}
\hline
\multirow{2}{*}{\textbf{Variant}} 
& \multirow{2}{*}{\textbf{EM}} 
& \multirow{2}{*}{\textbf{Context}} 
& \multicolumn{4}{c|}{\textbf{NDCG@k}} 
& \multicolumn{4}{c|}{\textbf{Recall@k}} 
& \multirow{2}{*}{\textbf{MRR}} \\
\cline{4-11}
& & 
& @1 & @3 & @5 & @10 
& @1 & @3 & @5 & @10 
& \\ 
\hline
TOPSIS labels + full features
& $\times$ & $\checkmark$ 
& 0.9423 & 0.9616 & 0.9724 & 0.9759 
& 0.2516 & 0.7401 & 0.9596 & 0.9986 
& 0.9981 \\

TOPSIS labels + candidate-only features 
& $\times$ & $\times$ 
& 0.7508 & 0.8478 & 0.8845 & 0.8989 
& 0.2426 & 0.6936 & 0.9120 & 0.9925 
& 0.9764 \\

EM labels + candidate-only features 
& $\checkmark$  & $\times$ 
& 0.8802 & 0.8933 & 0.9130 & 0.9317 
& 0.1902 & 0.5061 & 0.6815 & 0.8835 
& 0.9960 \\

EM labels + full features (Full Model) 
& $\checkmark$  & $\checkmark$  
& 0.9795 & 0.9725 & 0.9762 & 0.9811 
& 0.1931 & 0.5112 & 0.6848 & 0.8889 
& 0.9990 \\
\hline
\end{tabular}
\end{table*}

\section{Limitations}
\label{limitation}

Although the proposed framework demonstrates robust predictive and calibration performance, several limitations have been identified.
\emph{First}, the charging station choices are extracted from historical mobility traces rather than observed charging or EV--EV transaction records. Consequently, the data do not directly verify whether a stop corresponded to an actual charging or discharging event, which station was used, how much energy was transferred, how long the transaction lasted, or whether the interaction involved grid charging or EV--EV trading. This introduces uncertainty into the derived ground truth and may lead to discrepancies between inferred charging decisions and real operational behavior.
\emph{Second}, the station popularity is normalized relative to the average percentage rather than the absolute demand or supply, which may obscure real-world charging or discharging load variations across different regions and times.
\emph{Third}, while heuristic and EM-based soft labeling support label quality, both rely on underlying assumptions about decision rationality and feature relevance, which may not fully capture all behavioral or environmental factors influencing charging/discharging choices.
\emph{Finally}, the current analysis focuses on station-level recommendation without incorporating real-time changes to energy prices, which are critical for deployment in large-scale EV–EV trading systems.
Despite these limitations, the proposed ranking framework provides a strong methodological foundation for scalable, data-driven coordination in EV–EV energy trading.

\section{Conclusion and Future Works}
\label{conclusion}

This study presented a recommendation framework for predicting next charging nodes in EV–EV energy trading scenarios, explicitly addressing the uncertainty and multi-choice nature of charging decisions along EV journeys. Rather than modeling charging behavior as a single-label prediction problem, the proposed approach formulates next charging nodes selection as a learning-to-rank task, where candidate stations are ordered according to their relative suitability within each decision event.

The framework integrates fuzzy-weighted TOPSIS for role-aware heuristic suitability estimation, EM-based probabilistic smoothing to model decision ambiguity, and supervised learning-to-rank models to learn a generalized ranking policy. Experimental evaluation on the NextTrade-EV dataset demonstrated that gradient-boosted ranking models are well suited to this setting, with LightGBM (LambdaRank) consistently achieving the strongest performance across NDCG@k, Recall@k, and MRR metrics. Results further showed that probabilistic smoothing alone does not enhance heuristic rankings, but becomes effective when paired with rich EV-level and contextual features, enabling improved early-ranking accuracy and robust generalization to unseen decision events.

Beyond empirical performance, this study contributes to a reproducible and extensible evaluation pipeline for EV–EV charging node recommendation under uncertainty. By explicitly modeling role-dependent urgency, energy level dynamics, temporal context, and candidate-set characteristics, the framework captures realistic variability in EV–EV trading behavior without relying on observed station selection labels. This design supports decentralized coordination by producing ranked candidate sets rather than point predictions, which is more aligned with EV–EV matching and negotiation processes.

As EV participation and trading density continue to scale, our future research development will focus on improving model scalability and privacy through distributed and privacy-preserving learning paradigms. In addition, while the current formulation assumes stationary charging nodes as intermediaries, real-world deployments may involve mobile charging units (charging-as-a-service) or direct EV–EV energy exchange. Extending the framework to accommodate such dynamic trading configurations represents an important direction for further research.











\section*{Acknowledgments}
The authors acknowledge that this research is partly supported by the QUT Postgraduate Research Award (QUTPRA)
scholarship.

\section*{Data Availability}
The data that support the findings of this study are available from the corresponding author upon reasonable request.
\bibliographystyle{cas-model2-names}

\bibliography{cas-refs}



\end{document}